\pdfoutput=1
\documentclass[iop]{emulateapj}
\slugcomment{}

\shorttitle{New insights into the physical processes in the mid-infrared bubble N49}
\shortauthors{L.~K. Dewangan et al.}
\begin{document}

\title{New insights in the mid-infrared bubble N49 site: a clue of collision of filamentary molecular clouds}
\author{L.~K. Dewangan\altaffilmark{1}, D.~K. Ojha\altaffilmark{2}, and I. Zinchenko\altaffilmark{3}}
\email{lokeshd@prl.res.in}
\altaffiltext{1}{Physical Research Laboratory, Navrangpura, Ahmedabad - 380 009, India.}
\altaffiltext{2}{Department of Astronomy and Astrophysics, Tata Institute of Fundamental Research, Homi Bhabha Road, Mumbai 400 005, India.}
\altaffiltext{3}{Institute of Applied Physics of the Russian Academy of Sciences, 46 Ulyanov st., Nizhny Novgorod 603950, Russia.}
\begin{abstract}
We investigate the star formation processes operating in a mid-infrared bubble N49 site,
which harbors an O-type star in its interior, an ultracompact H\,{\sc ii} region, 
and a 6.7 GHz methanol maser at its edges. 
The $^{13}$CO line data reveal two velocity components (at velocity peaks $\sim$88 and $\sim$95 km s$^{-1}$) in the direction of the bubble. An elongated filamentary feature (length $>$15 pc) is investigated in each molecular cloud component, and the bubble is found at the interface of these two filamentary molecular clouds. 
The {\it Herschel} temperature map traces all these structures in a temperature range of $\sim$16--24 K. 
In the velocity space of $^{13}$CO, the two molecular clouds are separated by $\sim$7 km s$^{-1}$, and are interconnected by a lower intensity intermediate 
velocity emission (i.e. a broad bridge feature). A possible complementary molecular pair at [87, 88] km s$^{-1}$ and [95, 96] km s$^{-1}$ is also observed in the velocity channel maps. 
These observational signatures are in agreement with the outcomes of simulations 
of the cloud-cloud collision process. 
There are also noticeable embedded protostars and {\it Herschel} clumps distributed toward the filamentary features including the intersection zone of the two molecular clouds. 
In the bubble site, different early evolutionary stages of massive star formation are also present.
Together, these observational results suggest that in the bubble N49 site, the collision of the filamentary molecular clouds appears to be operated about 0.7 Myr ago, and may have triggered the formation of embedded protostars and massive stars. 
\end{abstract}
\section{Introduction}
\label{sec:intro}
Massive stars ($\geq$ 8 M$_{\odot}$) can inject large amounts of energy to the 
neighboring interstellar medium (ISM), hence these stars can 
trigger the birth of a new generation of stars including young massive star(s) \citep{deharveng10}. 
However, the formation mechanisms of massive stars and their feedback processes are still being debated \citep{zinnecker07,tan14}. 
In recent years, the theoretical and observational studies of the cloud-cloud collision (CCC) 
process have drawn considerable attention, which can produce massive OB stars and young stellar clusters at the junction of molecular clouds \citep[e.g.,][]{habe92,furukawa09,anathpindika10,ohama10,inoue13,takahira14,fukui14,fukui16,torii15,torii17,haworth15a,haworth15b,dewangan17a,dewangan17b}. \citet{torii17} suggested that the onset of the CCC process in a given star-forming region could be observationally inferred through the detection of a bridge feature 
connecting the two clouds in velocity space, the broad CO line wing in the intersection of the two clouds, 
and the complementary distribution of the two colliding clouds. 
However, such observational investigation is still limited in the literature \citep[e.g.][]{torii17}.

The mid-infrared (MIR) bubble, N49 \citep[{\it l} = 028$\degr$.827; {\it b} = $-$00$\degr$.229;][]{churchwell06} is a very well studied star-forming site containing an H\,{\sc ii} region \citep{watson08,anderson09,deharveng10,everett10,zavagno10,dirienzo12}. 
The N49 H\,{\sc ii} region is ionized by an O type star \citep{watson08,deharveng10,dirienzo12} and is situated at a distance of 5.07 kpc \citep{dirienzo12}. We have also adopted a distance of 5.07 kpc to the bubble N49 throughout the present work.
The bubble N49 is classified as a complete or closed ring with an average radius 
and thickness of 1$\farcm$32 (or 1.95 pc) and 0$\farcm$32 (or 0.45 pc), respectively \citep{churchwell06}.
\citet{dirienzo12} examined the $^{13}$CO line data and suggested the presence of two velocity components (at $\sim$87 and $\sim$95 km s$^{-1}$) in the direction of the bubble. 
Based on the radio recombination line observations, the velocity of the ionized gas in the N49 H\,{\sc ii} region was reported to be $\sim$90.6 km s$^{-1}$ \citep{anderson09}.
Using the APEX 870 $\mu$m dust continuum data, \citet{deharveng10} reported the detection of at least four massive clumps (M$_{clump}$ $\sim$ 190--2300 M$_{\odot}$) toward 
the infrared rim of the bubble (see Figure~18 in \citet{deharveng10} and also Figure~1 in \citet{zavagno10}).  
An ultracompact (UC) H\,{\sc ii} region and a 6.7 GHz methanol maser emission (MME) \citep[velocity range $\sim$79.4--92.7 km s$^{-1}$;][]{walsh98,szymczak12} are also detected toward the dust condensations, which are seen at the edges of the bubble \citep[e.g.][]{deharveng10,zavagno10,dirienzo12}. 
The bubble N49 has also been considered as a candidate of a wind-blown bubble \citep{everett10}. 
Using the multi-wavelength data, previous studies suggested that the N49 H\,{\sc ii} region is interacting with its surrounding molecular cloud, and has been cited as a possible site of triggered star formation \citep{watson08,anderson09,zavagno10,deharveng10,dirienzo12}. 

Despite the availability of several observational data sets, 
the knowledge of the physical environments over larger spatial scale around the bubble is still unknown.
Furthermore, the study of an interaction between molecular cloud components is yet to be performed in the bubble N49 site.
To study the physical environment and star formation processes around the bubble N49, 
we revisit the bubble using multi-wavelength data covering from the radio to near-infrared (NIR) wavelengths. Such analysis offers an opportunity to examine the distribution of dust temperature, column density, extinction, ionized emission, kinematics of molecular gas, and young stellar objects (YSOs). 

The paper is arranged in the following way. 
The details of the adopted data sets are described in Section~\ref{sec:obser}. 
Section~\ref{sec:data} gives the outcomes concerning to the physical environment and point-like sources.  
In Section~\ref{sec:disc}, we present the possible star formation processes ongoing 
in our selected target region. Finally, Section~\ref{sec:conc} summarizes the main results.
\section{Data sets and analysis}
\label{sec:obser}
In this paper, we have chosen a field of $\sim$0$\degr$.42 
$\times$ 0$\degr$.42 ($\sim$37.2 pc $\times$ 37.2 pc; centered at $l$ = 28$\degr$.844; $b$ = $-$0$\degr$.220) around the MIR bubble N49 site. 
In the following, we provide a brief description of the adopted multi-wavelength data. 
\subsection{Radio Centimeter Continuum Map}
Radio continuum map at 20 cm was obtained from the VLA 
Multi-Array Galactic Plane Imaging Survey \citep[MAGPIS;][]{helfand06}. 
The MAGPIS 20 cm map has a 6\farcs2 $\times$ 5\farcs4 beam size and a pixel scale of 2$\arcsec$/pixel. 
\subsection{$^{13}$CO (J=1$-$0) Line Data}
In order to examine the molecular gas associated with the selected target, 
the Galactic Ring Survey \citep[GRS;][]{jackson06} $^{13}$CO (J=1$-$0) line data were adopted. 
The GRS line data have a velocity resolution of 0.21~km\,s$^{-1}$, an angular resolution 
of 45$\arcsec$ with 22$\arcsec$ sampling, a main beam efficiency ($\eta_{\rm mb}$) of $\sim$0.48, 
a velocity coverage of $-$5 to 135~km~s$^{-1}$, and a typical rms sensitivity (1$\sigma$)
of $\approx0.13$~K \citep{jackson06}.  
\subsection{Far-infrared and Sub-millimeter Data}
We examined the far-infrared (FIR) and sub-millimeter (mm) images downloaded from the {\it Herschel} Space Observatory \citep{pilbratt10,poglitsch10,griffin10,degraauw10} data archives. 
Level2$_{-}$5 processed 160--500 $\mu$m images were retrieved through the {\it Herschel} Interactive Processing Environment \citep[HIPE,][]{ott10}. 
The beam sizes of the {\it Herschel} images are 5$\farcs$8, 12$\arcsec$, 18$\arcsec$, 25$\arcsec$, and 37$\arcsec$ for 70, 160, 250, 350, and 500 $\mu$m, respectively \citep{poglitsch10,griffin10}. 
The plate scales of 70, 160, 250, 350, and 500 $\mu$m images are 3$''$.2, 3$''$.2, 6$''$, 10$''$, and 14$''$ pixel$^{-1}$, respectively.  
The {\it Herschel} images at 250--500 $\mu$m are calibrated in units of surface brightness, MJy sr$^{-1}$, while the units of images at 70--160 $\mu$m 
are Jy pixel$^{-1}$. 

The sub-mm continuum map at 870 $\mu$m (beam size $\sim$19$\farcs$2) was also retrieved from the APEX Telescope Large Area Survey of the Galaxy \citep[ATLASGAL;][]{schuller09}. 
\subsection{{\it Spitzer} and {\it WISE} Data}
The photometric images and magnitudes of point sources at 3.6--8.0 $\mu$m were downloaded 
from the {\it Spitzer} Galactic Legacy Infrared Mid-Plane Survey Extraordinaire \citep[GLIMPSE;][]{benjamin03} survey (resolution $\sim$2$\arcsec$). 
In this work, we used the GLIMPSE-I Spring '07 highly reliable photometric catalog. 

We also utilized the Wide Field Infrared Survey Explorer (WISE\footnote[1]{WISE is a joint project of the
University of California and the JPL, Caltech, funded by the NASA}; \citet{wright10}) image at 12 $\mu$m (spatial resolution $\sim$6$\arcsec$) and the {\it Spitzer} MIPS Inner Galactic Plane Survey \citep[MIPSGAL;][]{carey05} 24 $\mu$m image (spatial resolution $\sim$6$\arcsec$). Furthermore, the photometric magnitudes of point sources at MIPSGAL 24 $\mu$m \citep[from][]{gutermuth15} were also collected. 
\section{Results}
\label{sec:data}
\subsection{MIR bubble N49 and filamentary features}
\label{subsec:u1}
\begin{figure*}
\epsscale{0.6}
\plotone{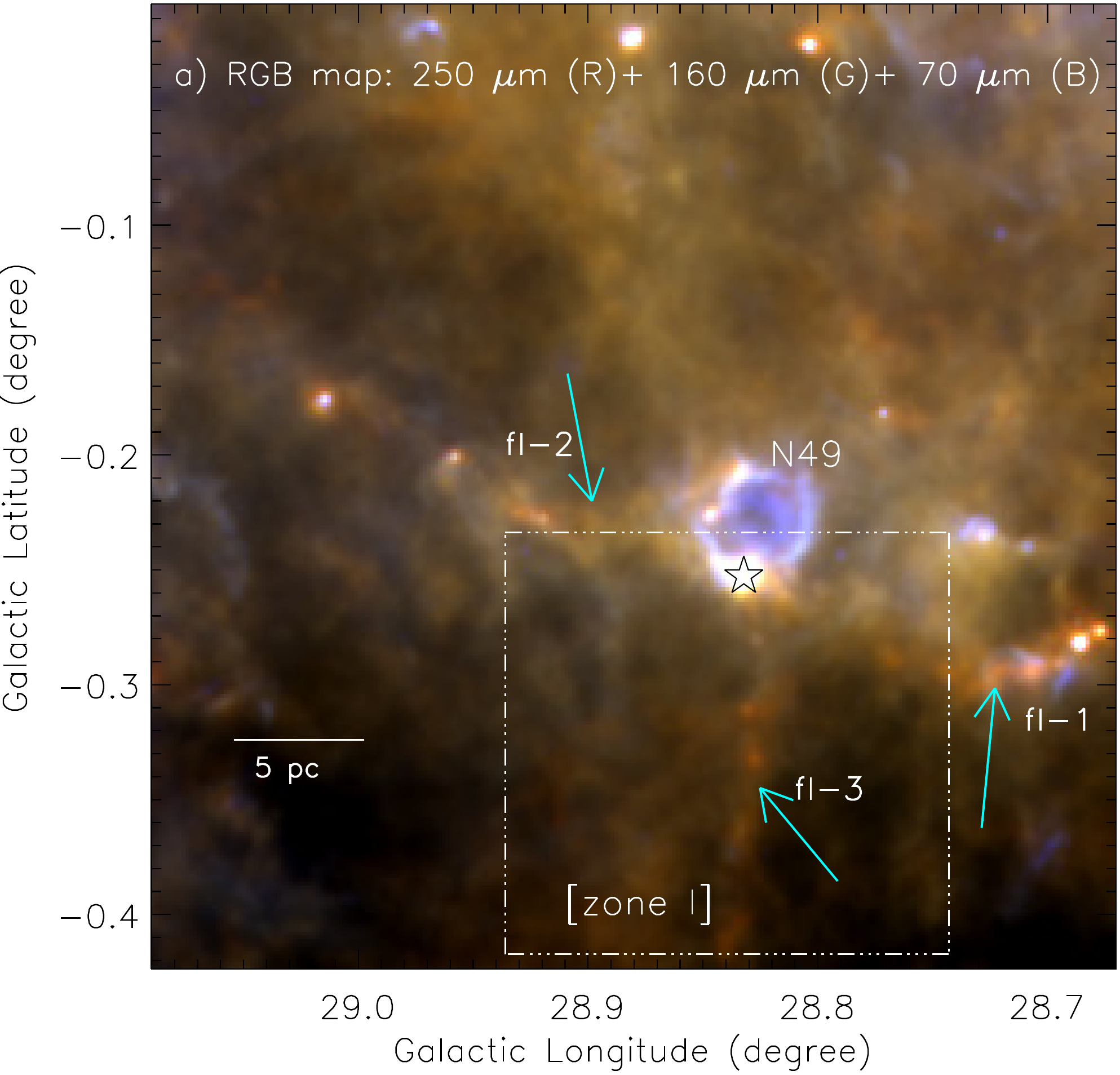}
\epsscale{0.52}
\plotone{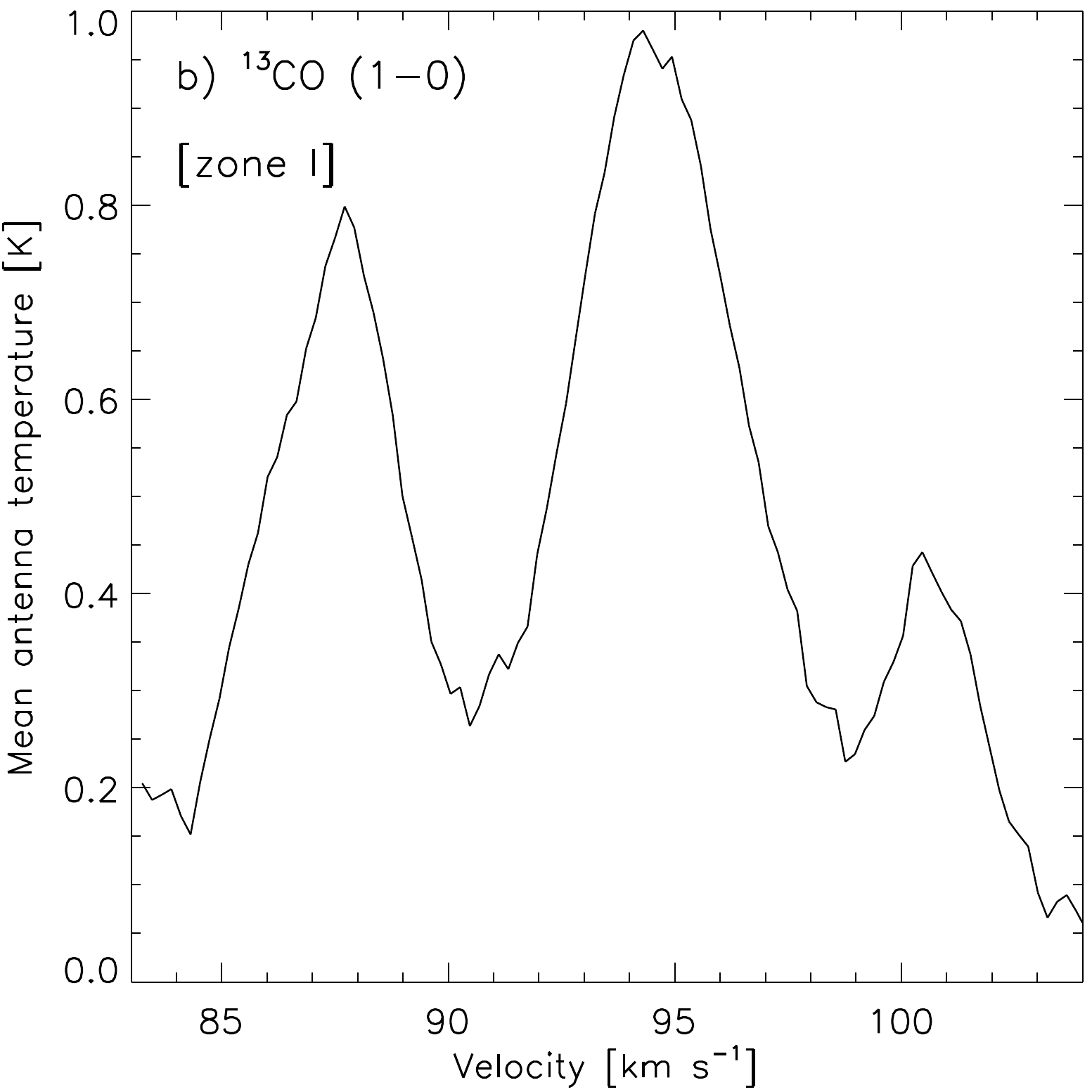}
\caption{\scriptsize a) Large-scale view of the MIR bubble N49 (size of the selected field $\sim$0$\degr$.42 
$\times$ 0$\degr$.42 ($\sim$37.2 pc $\times$ 37.2 pc); central coordinates: $l$ = 28$\degr$.844; $b$ = $-$0$\degr$.220). 
a) A three color-composite map ({\it Herschel} 250 $\mu$m (red), 160 $\mu$m (green), and 70 $\mu$m (blue) images in log scale). The MIR bubble N49 is prominently seen in the composite map. 
Three arrows highlight the embedded filaments. A position of the Class~II 6.7 GHz methanol 
maser \citep[from][]{szymczak12} is shown by a star, which is traced in a velocity 
range of 79.4 to 92.7 km s$^{-1}$. A scale bar corresponding to 5 pc is shown in the bottom left corner. b) The GRS $^{13}$CO (1-0) spectrum in the direction of a small field (i.e. zone I; see highlighted box in Figure~\ref{fig1}a).
The spectrum reveals three velocity components in the direction of zone I and is obtained by averaging the area.}
\label{fig1}
\end{figure*}

\begin{figure*}
\epsscale{1}
\plotone{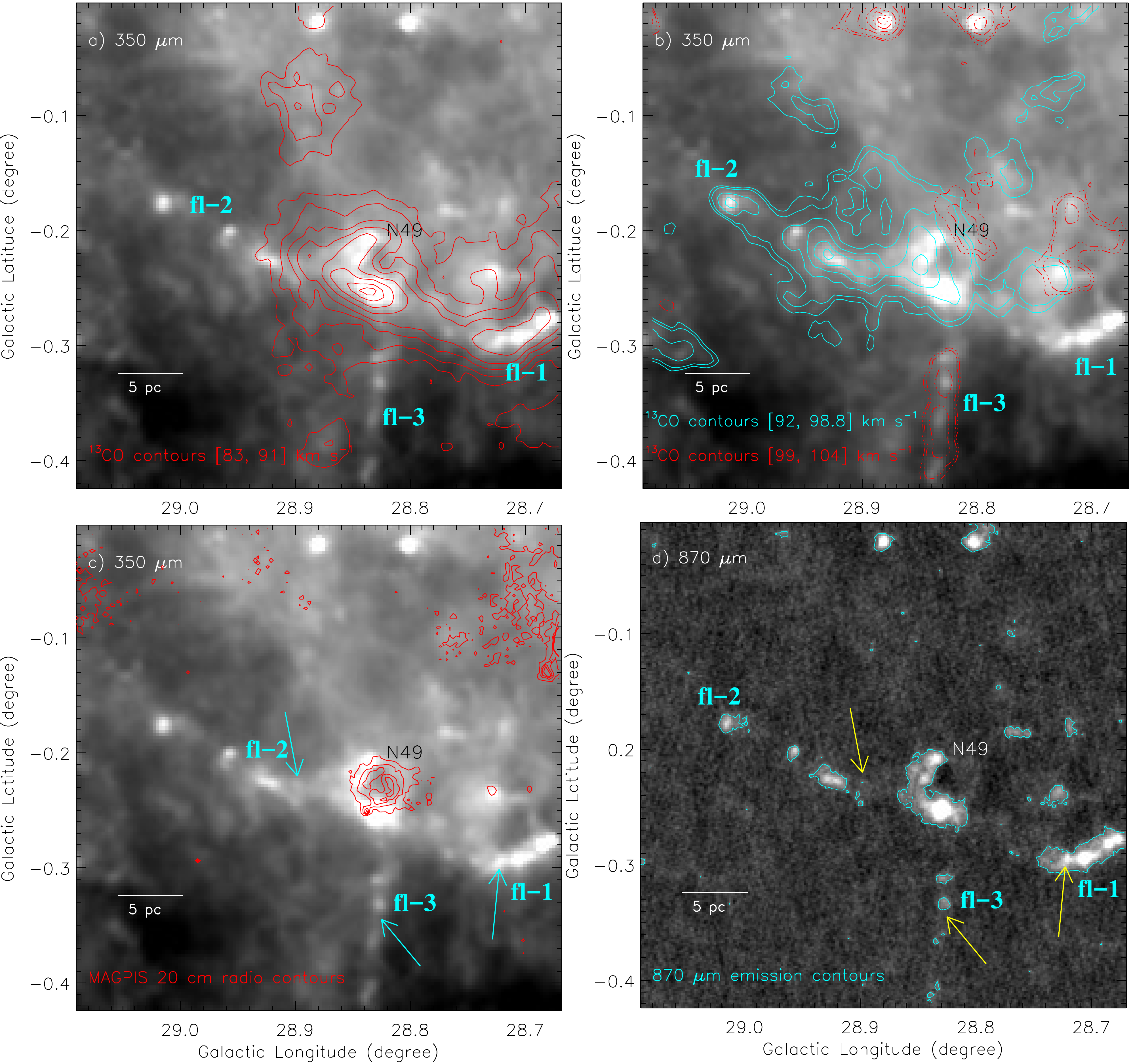}
\caption{\scriptsize a) Overlay of the molecular $^{13}$CO gas on the {\it Herschel} 350 $\mu$m image. The $^{13}$CO integrated velocity range is from 83 to 91 km s$^{-1}$. 
The $^{13}$CO contours are 40.847 K km s$^{-1}$ $\times$ (0.12, 0.2, 0.3, 0.4, 0.55, 0.7, 0.85, 0.95).
b) The $^{13}$CO emissions are superimposed on the {\it Herschel} 350 $\mu$m image.
The cyan and red contours represent $^{13}$CO integrated over two different velocity ranges highlighted in the image. The CO contours (in cyan) are 38.357 K km s$^{-1}$ $\times$ (0.35, 0.4, 0.55, 0.7, 0.85, 0.95), while the levels of red contours are 21.468 K km s$^{-1}$ $\times$ (0.35, 0.4, 0.55, 0.7, 0.85, 0.95).
c) Overlay of the MAGPIS 20 cm contours on the {\it Herschel} 350 $\mu$m image. 
The contours (in red) are shown with levels 
of 0.0243 Jy/beam $\times$ (0.06, 0.1, 0.14, 0.18). 
d) Overlay of the ATLASGAL 870 $\mu$m contour on the ATLASGAL 870 $\mu$m image.
A contour (in cyan) is shown with a level of 0.163 Jy/beam. 
In each panel, a scale bar corresponding to 5 pc is shown in the bottom left corner.
In the panels c and d, three arrows highlight the embedded filamentary features.}
\label{fig2}
\end{figure*}

In this section, we present multi-wavelength data to explore the physical environments over larger spatial scale around the bubble N49.
Figure~\ref{fig1}a shows a color-composite map obtained using the {\it Herschel} images 
(i.e. 250 $\mu$m (red), 160 $\mu$m (green), and 70 $\mu$m (blue)). 
The MIR bubble N49 is prominently seen in the composite map 
within a spatial area of 6.5 pc $\times$ 6.5 pc, and the {\it Herschel} images also 
reveal embedded filamentary features (i.e. fl-1, fl-2, and fl-3) in our selected field (see arrows in Figure~\ref{fig1}a).
We also find that the bubble appears at the junction of filamentary features. 
However, one cannot confirm the physical association between the bubble and filamentary features without knowledge of velocities of molecular gas.
In Figure~\ref{fig1}b, we present the observed $^{13}$CO (J=1--0) profile 
in the direction of ``zone I" (see a highlighted box in Figure~\ref{fig1}a), which encompasses spatially some parts of the bubble and the filamentary features.
The spectrum is obtained by averaging the ``zone I" area, 
and reveals the presence of at least three velocity components (at peaks around 88, 95, and 100 km s$^{-1}$) along the line of sight. 
Based on the $^{13}$CO spectrum, in Figures~\ref{fig2}a and~\ref{fig2}b, we present the overlay of the $^{13}$CO emissions on the {\it Herschel} 350 $\mu$m image. 
In Figure~\ref{fig2}a, the $^{13}$CO gas is integrated over a velocity range of 83--91 km s$^{-1}$, and a majority of molecular gas are found toward 
the bubble and the filamentary feature ``fl-1". 
The distribution of molecular gases linked with two other molecular cloud components is presented in Figure~\ref{fig2}b. 
The molecular cloud linked with the filamentary feature ``fl-2" is traced in a velocity range of 92--98.8 km s$^{-1}$, while the molecular cloud associated with the filamentary feature ``fl-3" is depicted in a velocity range of 99--104 km s$^{-1}$. 

\begin{figure*}
\epsscale{1}
\plotone{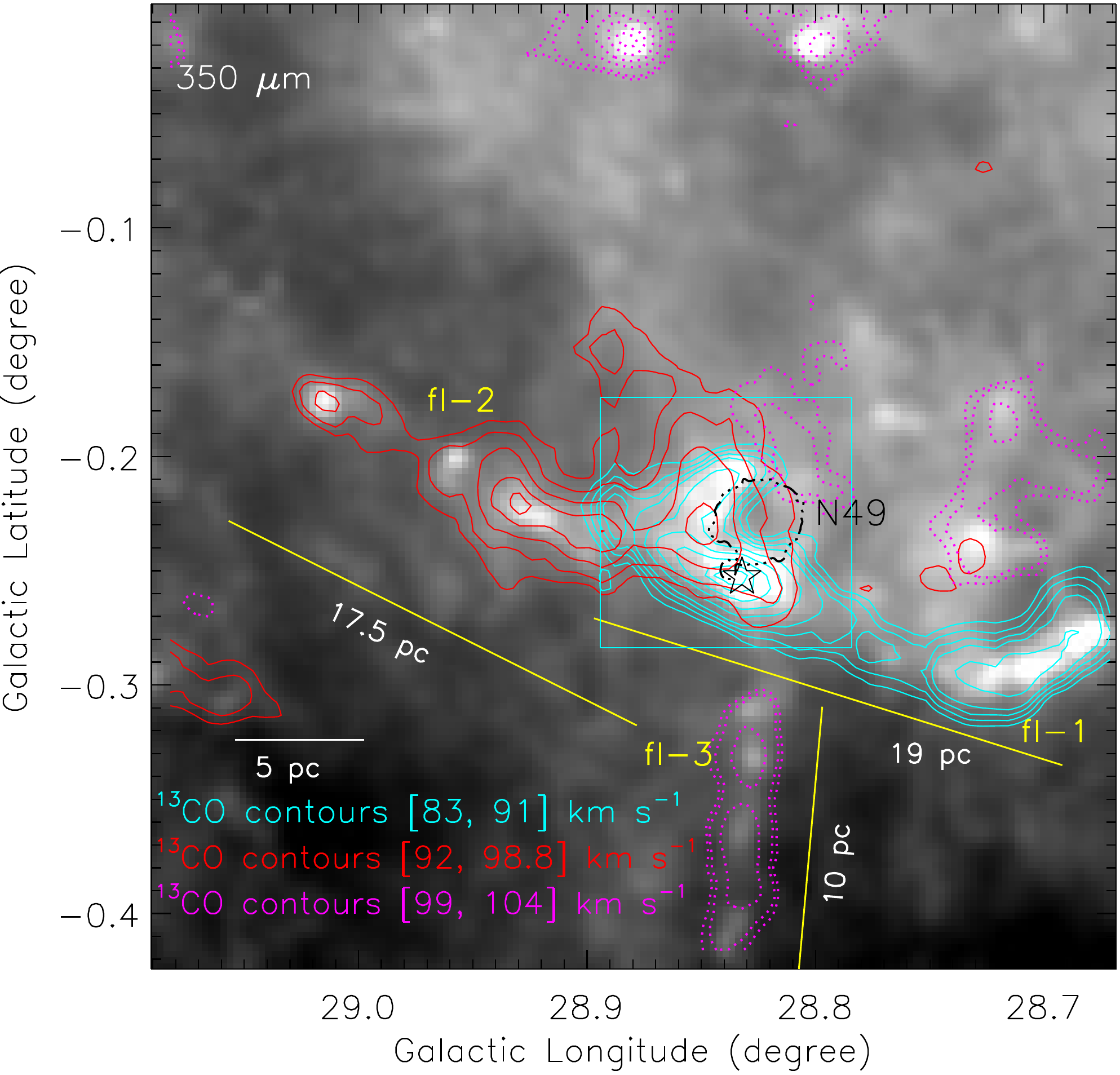}
\caption{\scriptsize Overlay of the molecular clouds linked with the filamentary features on the {\it Herschel} 350 $\mu$m image. The $^{13}$CO emissions integrated over three different velocity ranges are presented and the velocity ranges are also given in the figure (also see Figures~\ref{fig2}a and~\ref{fig2}b). 
The $^{13}$CO contours (in red) are 38.357 K km s$^{-1}$ $\times$ (0.45, 0.55, 0.7, 0.85, 0.95), while the levels of cyan contours are 40.847 K km s$^{-1}$ $\times$ (0.35, 0.4, 0.45, 0.5, 0.6, 0.7, 0.8, 0.9, 0.95). 
The $^{13}$CO contours (in magenta) are similar to those shown in Figures~\ref{fig2}b. A broken contour at 20 cm (in black) represents the location of the bubble N49, and the contour level is 0.0024 Jy/beam. A position of the Class~II 6.7 GHz methanol maser \citep[from][]{szymczak12} is shown by a black star. 
The solid box (in cyan) encompasses the area shown in Figures~\ref{fig3}a-l and~\ref{fig4}a. A scale bar corresponding to 5 pc  is shown in the bottom left corner.}
\label{fig2xx}
\end{figure*}

Figure~\ref{fig2}c displays the {\it Herschel} 350 $\mu$m image overlaid with the MAGPIS 20 cm emission. The ionized emission traced in the MAGPIS map is exclusively seen toward the bubble N49.
In Figure~\ref{fig2}d, the ATLASGAL 870 $\mu$m continuum map is also superimposed with the 870 $\mu$m emission contour, indicating the presence of several condensations toward the filamentary features and the edges of the bubble. 
In Figures~\ref{fig2}c and~\ref{fig2}d, despite the difference in spatial 
resolution, one can infer that the emission traced in the {\it Herschel} 350 $\mu$m is found to be more prominent compared to the emission detected in the 870 $\mu$m continuum map. 
It has also been reported that space-based {\it Herschel} observations could be 
considered as almost no loss of large-scale emission with respect to the ground-based APEX dust continuum observations \citep[e.g.][]{liu17}.
In our selected target field, based on the distribution of molecular gas, 
Figure~\ref{fig2xx} spatially delineates different elongated filamentary molecular clouds (lengths $\sim$10--19 pc; average widths $\sim$2 pc).

\begin{figure*}
\epsscale{0.86}
\plotone{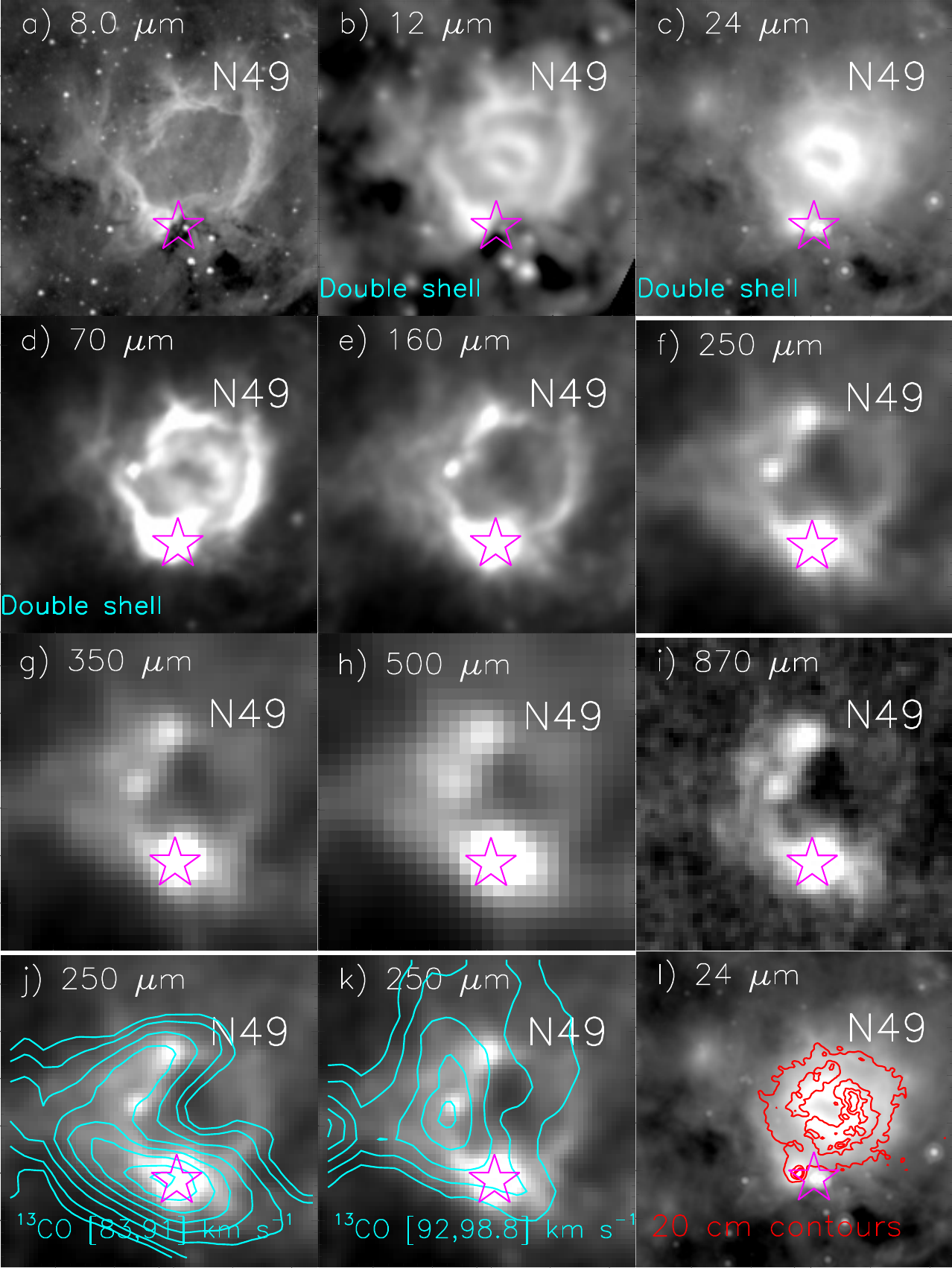}
\caption{\scriptsize A multi-wavelength view of the bubble N49. The images are shown at different wavelengths, which are highlighted in the panels. In the panels ``j" and ``k", the $^{13}$CO emissions are superimposed on the {\it Herschel} 250 $\mu$m image. The $^{13}$CO integrated velocity ranges are also shown in these panels. 
In the panel ``j", the $^{13}$CO contours are 40.847 K km s$^{-1}$ $\times$ 
(0.35, 0.4, 0.5, 0.6, 0.7, 0.8, 0.9, 0.98). In the panel ``k", the $^{13}$CO contours are 38.357 K km s$^{-1}$ $\times$ (0.6, 0.7, 0.8, 0.9, 0.98). In the panel ``l", the MAGPIS 20 cm emission is superimposed on the {\it Spitzer} 24 $\mu$m image. The MAGPIS contours (in red) are shown with levels 
of 0.0018, 0.0035, 0.0045, and 0.005 Jy/beam. 
In each panel, a position of the Class~II 6.7 GHz methanol maser is shown by a star.}
\label{fig3}
\end{figure*}

Figure~\ref{fig3} shows a zoomed-in view of the bubble N49 using multi-wavelength images (e.g. {\it Spitzer} 8--24 $\mu$m, WISE 12 $\mu$m, {\it Herschel} 70--500 $\mu$m, 
ATLASGAL 870 $\mu$m, GRS $^{13}$CO, and MAGPIS 20 cm). These images reveal a complete or closed ring morphology, containing the ionized emission in the bubble interior \citep[e.g.][]{watson08,zavagno10}. As mentioned before, the N49 H\,{\sc ii} region is powered by an O type star \citep{watson08,deharveng10,dirienzo12}. Furthermore, a double shell-like structure is also observed in the 12-70 $\mu$m and 20 cm maps \citep[e.g.][]{watson08}. 
The 6.7 GHz MME and the UCH\,{\sc ii} region are seen at the edges of the bubble. 
The UCH\,{\sc ii} region was reported to be ionized by a B0V star \citep[e.g.][]{deharveng10}. 
In the panels ``j" and ``k", one can also find the presence of two molecular cloud components in the direction of the bubble \citep[e.g.][]{dirienzo12}.
In Figure~\ref{fig4}a, we present a color-composite map produced using the MIR and FIR images 
(i.e. 70 $\mu$m (red), 24 $\mu$m (green), and 12 $\mu$m (blue)). The composite map is also overlaid with the MAGPIS 20 cm continuum emission, depicting the double shell-like structure. 
In the composite map, we have also highlighted the previously known UCH\,{\sc ii} region 
and two embedded YSOs (i.e. YSO \#1 and YSO\#3; see Figure~1 in \citet{zavagno10}). 
Interestingly, the position of the 6.7 GHz MME spatially coincides with the position of the YSO\#3 that can be considered as an infrared counterpart (IRc) of the 6.7 GHz MME. No radio cm emission is detected toward the YSO\#3.
Considering the 6.7 GHz MME as a reliable tracer of a massive YSO (MYSO) \citep[e.g.][]{walsh98,urquhart13}, the YSO\#3 could be a MYSO candidate at its early formation
stage prior to the UCH\,{\sc ii} phase. 
Based on the high resolution 6.7 GHz MME observations, \citet{cyganowski09} proposed the presence of a rotating disk associated with YSO\#3. 

Together, the bubble N49 is a very promising site, where different early evolutionary stages of massive star formation are present.
\subsection{Kinematics of molecular gas}
\label{sec:coem} 
\begin{figure*}
\epsscale{0.58}
\plotone{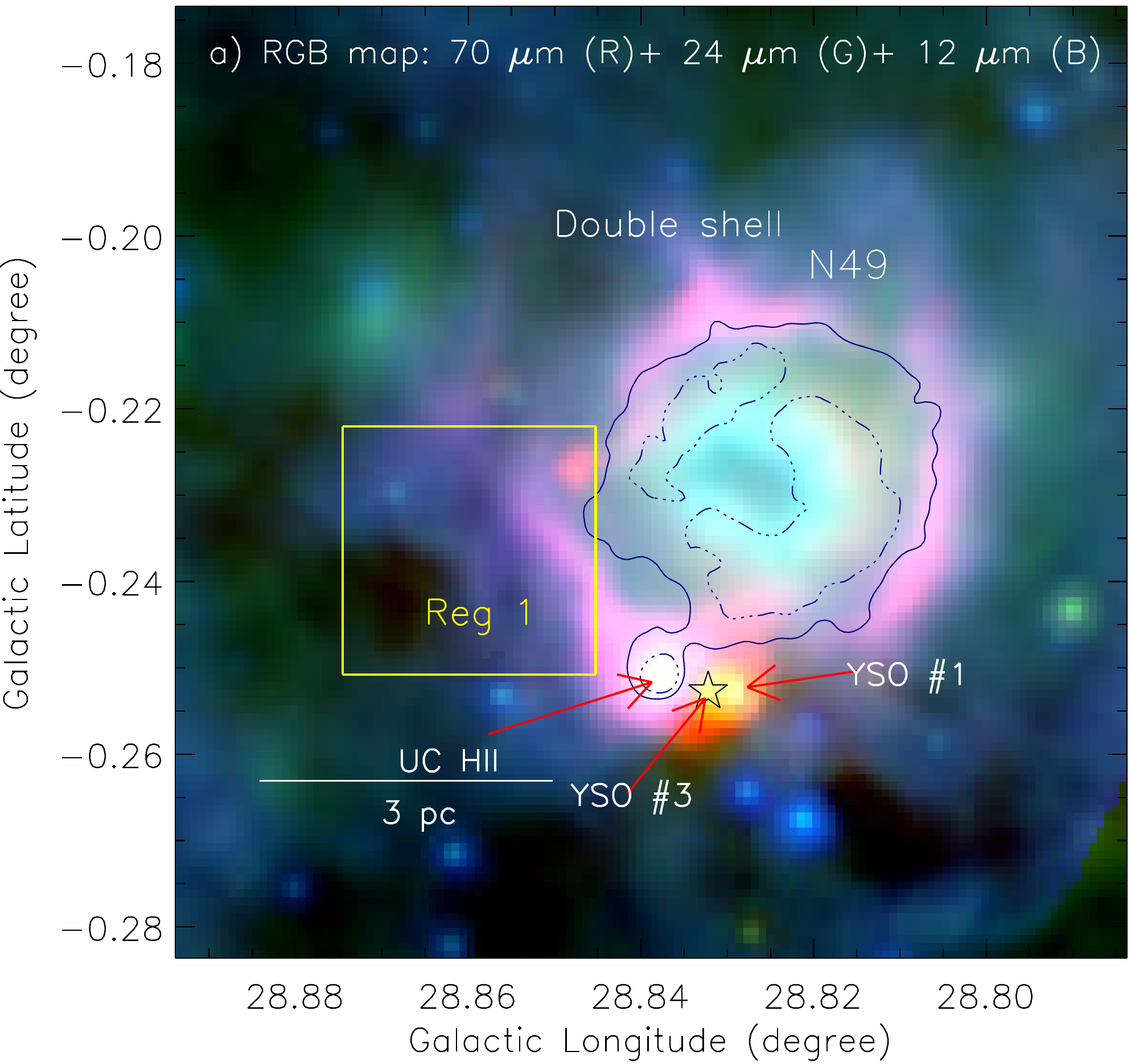}
\epsscale{0.58}
\plotone{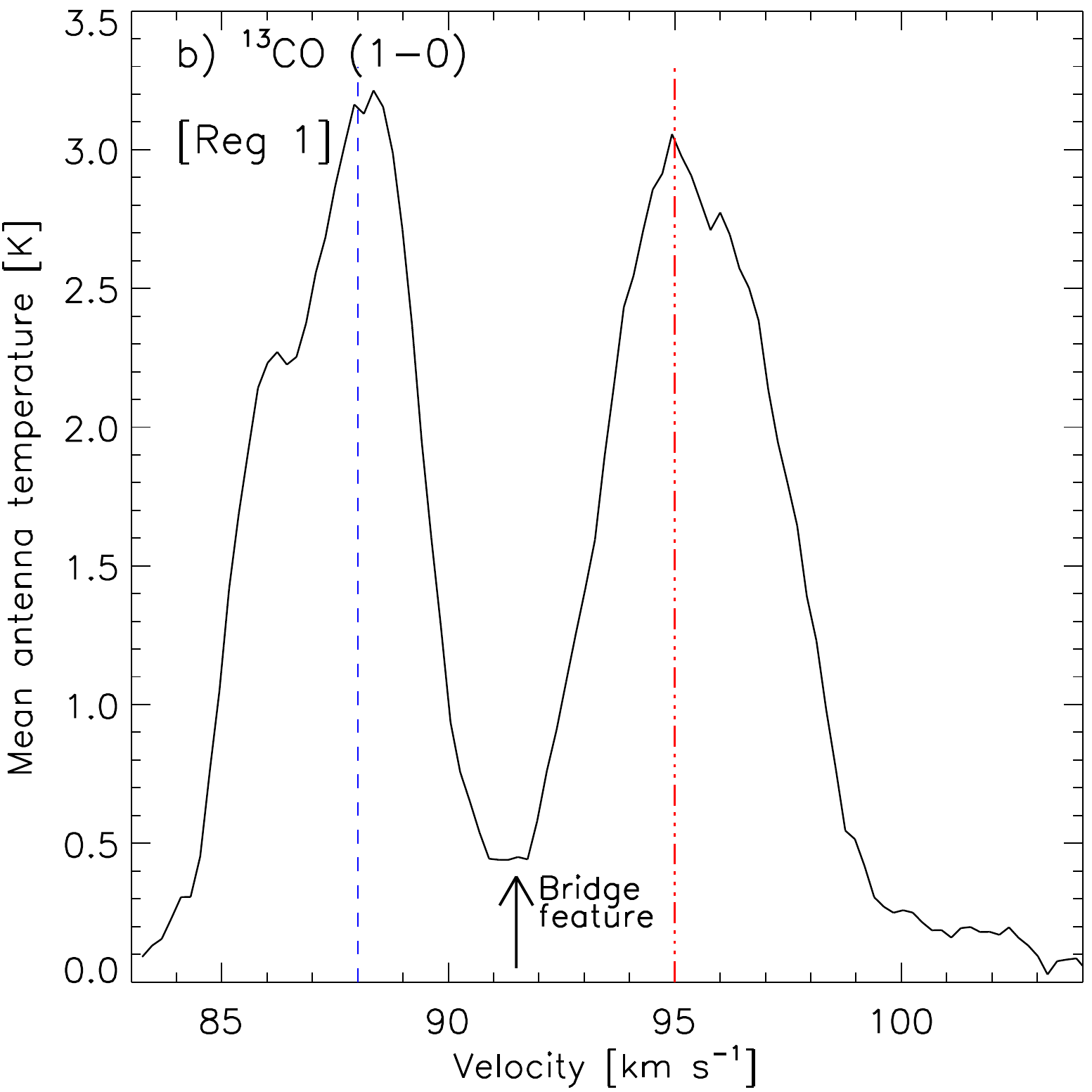}
\caption{\scriptsize Three-color composite image ({\it red}, {\it Herschel} 70 $\mu$m; {\it green}, {\it Spitzer} 24 $\mu$m; 
{\it blue},{\it WISE} 12 $\mu$m; in log scale) of the bubble N49, which clearly traces a double shell structure. The contours at 20 cm (in navy) are also overlaid on the composite image, and the contour levels are 0.0024 and 0.00353 Jy/beam. 
A position of the Class~II 6.7 GHz methanol maser is shown by a star. Previously three known sources (i.e. "UC HII", ``YSO \#1", and ``YSO \#3") are marked and highlighted by arrows in the figure \citep[also see Figure~1 in][]{zavagno10}. 
A scale bar corresponding to 3 pc is shown in the bottom left corner. 
b) The GRS $^{13}$CO (1-0) spectrum in the direction of a small field (i.e. Reg 1; see highlighted box in Figure~\ref{fig4}a). 
The spectrum is obtained by averaging the area, Reg 1.  Two velocity peaks are highlighted by broken lines. An almost flattened profile is found between two velocity peaks (i.e. a bridge feature).}
\label{fig4}
\end{figure*}

In this section, we present a kinematic analysis of the molecular gas in our selected field.
In Figure~\ref{fig4}b, we show the observed $^{13}$CO (J=1--0) spectrum in the direction of ``Reg 1" (see a solid box in Figure~\ref{fig4}a).
The spectrum is computed by averaging the area ``Reg 1" marked in Figure~\ref{fig4}a.
In the spectrum, an almost flattened profile is observed between two velocity peaks (or molecular cloud components), which can be referred to as a bridge feature at the intermediate velocity range.
This particular outcome indicates a signature of collisions between molecular clouds \citep{takahira14,haworth15a,haworth15b,torii17,bisbas17}. In other words, it also suggests a mutual interaction of clouds \citep[e.g.][]{bisbas17}.

\begin{figure*}
\epsscale{0.85}
\plotone{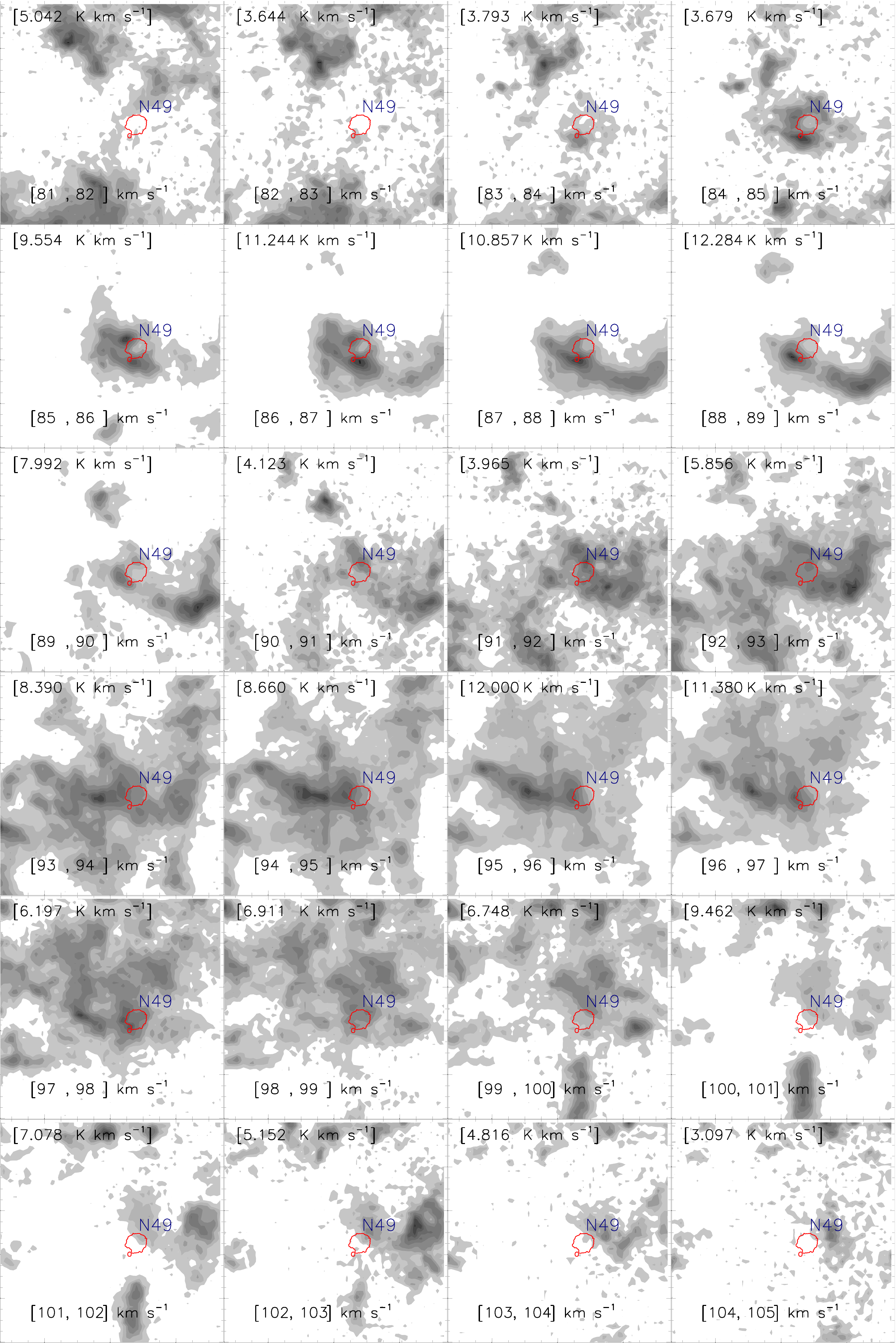}
\caption{\scriptsize The $^{13}$CO(J =1$-$0) velocity channel contour maps.
The molecular emission is integrated over a velocity interval, which is given in each panel (in km s$^{-1}$). 
The contour levels are 10, 20, 30, 40, 55, 70, 80, 90, and 99\% of the peak value (in K km s$^{-1}$), 
which is also given in each panel. In each panel, a contour at 20 cm (in red) represents the location of the bubble N49, 
and the contour level is 0.0024 Jy/beam.}
\label{fig5}
\end{figure*}

\begin{figure*}
\epsscale{1}
\plotone{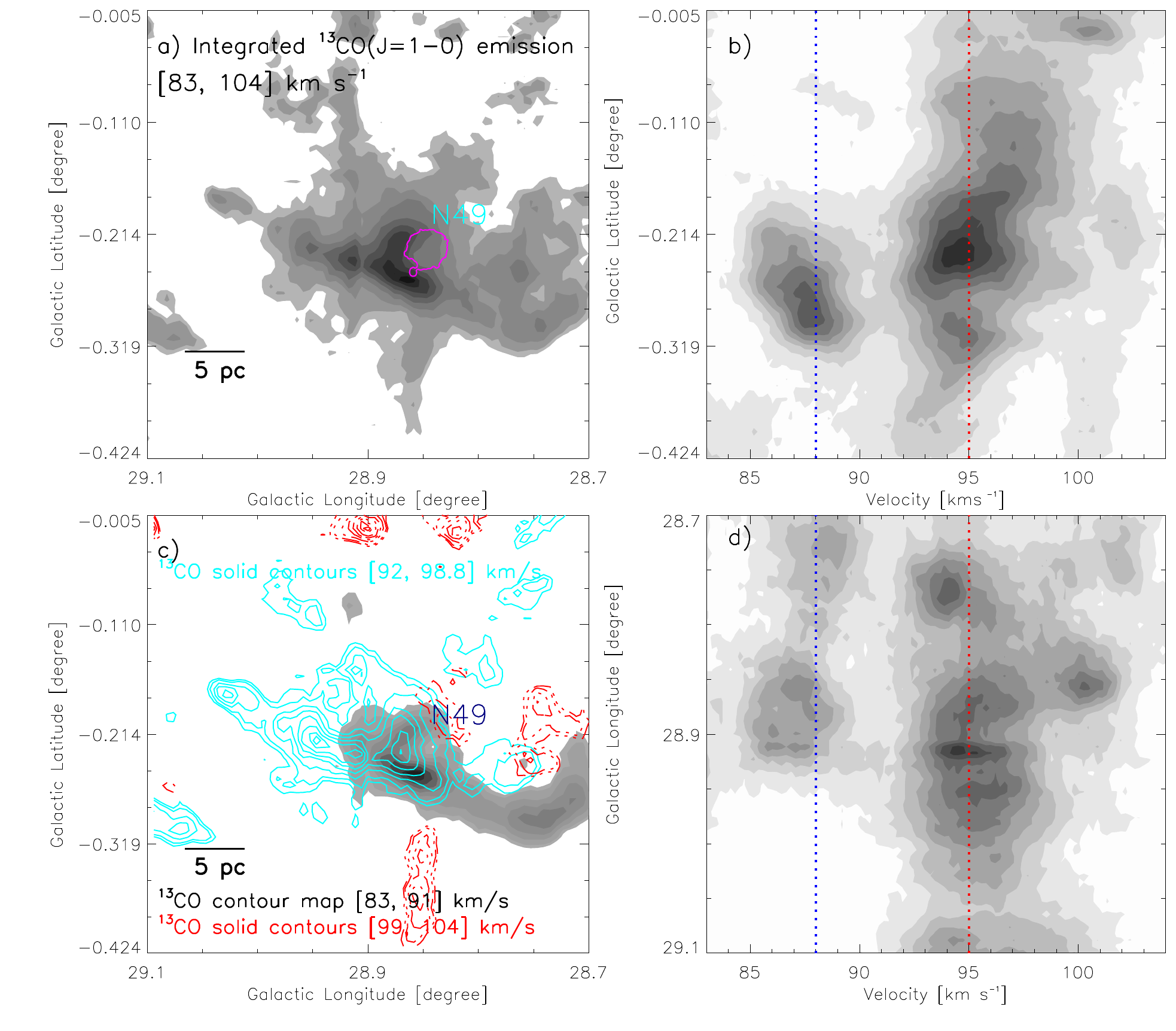}
\caption{\scriptsize 
a) Integrated intensity map of $^{13}$CO (J = 1-0) from 83 to 104 km s$^{-1}$. 
The contour levels are 25, 30, 40, 50, 60, 70, 80, 90, and 95\% of the 
peak value (i.e. 71.408 K km s$^{-1}$). A contour at 20 cm (in magenta) represents the location of the bubble N49, and the contour level is 0.0024 Jy/beam. 
b) Latitude-velocity map of $^{13}$CO. 
The $^{13}$CO emission is integrated over the longitude from 28$\degr$.67 to 29$\degr$.09. 
c) The $^{13}$CO emissions integrated over three different velocity ranges are presented, and the velocity ranges are also given in the panel (also see Figures~\ref{fig2}a and~\ref{fig2}b). 
d) Longitude-velocity map of $^{13}$CO. 
The $^{13}$CO emission is integrated over the latitude from $-$0$\degr$.424 to $-$0$\degr$.005. 
In both the left panels (i.e. Figures~\ref{fig6}a and~\ref{fig6}c), the scale bar corresponding to 5 pc is shown in the bottom left corner. 
Two dotted lines are also shown in both the right panels (i.e. Figures~\ref{fig6}b and~\ref{fig6}d), and are similar to those shown in Figure~\ref{fig4}b.}
\label{fig6}
\end{figure*}

In Figure~\ref{fig5}, we display the integrated GRS $^{13}$CO (J=1$-$0) velocity channel maps (starting from 81 km s$^{-1}$ at intervals of 1 km s$^{-1}$), tracing three molecular components along the line of sight (also see Figure~\ref{fig2xx}). 
To further examine the molecular gas distribution in the direction of our selected target field, 
in Figure~\ref{fig6}, we show the integrated $^{13}$CO intensity map and the position-velocity maps.  The integrated GRS $^{13}$CO intensity map is shown in Figure~\ref{fig6}a, where 
the molecular emission is integrated over 83 to 104 km s$^{-1}$.
The Galactic position-velocity diagrams of the $^{13}$CO emission also reveal three velocity components and the noticeable velocity spread (see Figures~\ref{fig6}b and~\ref{fig6}d). 
In the velocity space, we find a red-shifted peak (at $\sim$95 km s$^{-1}$) and a blue-shifted peak (at $\sim$88 km s$^{-1}$) that are interconnected by a lower intensity intermediate velocity emission, suggesting the presence of a broad bridge feature (also see Figure~\ref{fig4}b). 
Figure~\ref{fig6}c shows the spatial distribution of three molecular components, similar to those shown in Figures~\ref{fig2}a and~\ref{fig2}b.

Together, molecular line data confirm that the bubble N49 is found in the intersection of two molecular clouds. 
The analysis of $^{13}$CO data also gives an observational clue 
of the signature of the interaction between molecular cloud components in the bubble site.
The implication of these outcomes is presented in more detail in the discussion Section~\ref{sec:disc}.
\subsection{{\it Herschel} temperature and column density maps}
\label{subsec:temp}
In this section, we present {\it Herschel} temperature and column density maps of the bubble N49. Following the methods described in \citet{mallick15}, these maps are produced from a 
pixel-by-pixel spectral energy distribution (SED) fit with a modified blackbody to the cold dust emission at {\it Herschel} 160--500 $\mu$m \citep[also see][]{dewangan15}. 
In the following, a brief step-by-step explanation of the adopted procedures is provided. 

Before the SED fitting process, using the task ``Convert Image Unit" available in the HIPE software, 
we converted the surface brightness unit of 250--500 $\mu$m images to Jy pixel$^{-1}$, same as the unit of 160 $\mu$m image. Next, using the plug-in ``Photometric Convolution" available in the HIPE software, the 160--350 $\mu$m images were convolved to the angular resolution of 
the 500 $\mu$m image ($\sim$37$\arcsec$), and then regridded on a 14$\arcsec$ raster.
We then computed a background flux level. 
The sky background flux level was estimated to be 0.255, 0.708, 1.395, and $-$0.234 Jy pixel$^{-1}$ for the 500, 350, 250, and 
160 $\mu$m images (size of the selected featureless dark region $\sim$10$\farcm$2 $\times$ 9$\farcm$8; 
centered at:  $l$ = 27$\degr$.735; $b$ = $-$0$\degr$.681), respectively. The negative flux value at 160 $\mu$m is found due to the arbitrary scaling of the {\it Herschel} 160 $\mu$m image.

\begin{figure*}
\epsscale{0.57}
\plotone{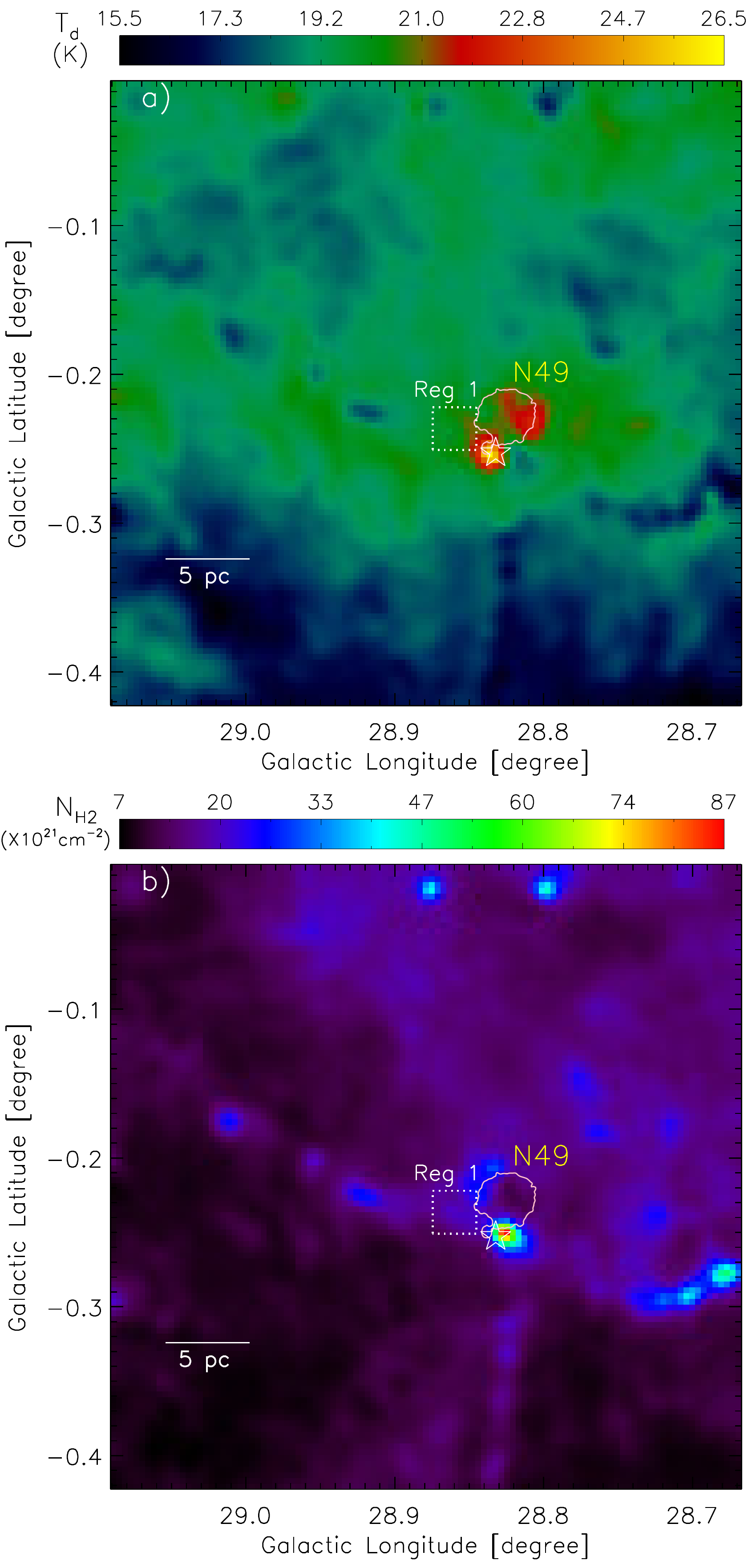}
\caption{\scriptsize a) {\it Herschel} temperature map. 
b) {\it Herschel} column density ($N(\mathrm H_2)$) map.
One can also compute the extinction value with $A_V=1.07 \times 10^{-21}~N(\mathrm H_2)$ \citep{bohlin78}. A position of the Class~II 6.7 GHz methanol maser is shown by a star in each panel. In both the panels, a small dotted square indicates the area, Reg1, which is similar to those shown in Figure~\ref{fig4}a. In each panel, a contour at 20 cm (in pink) represents the location of the bubble N49, and the contour level is 0.0024 Jy/beam. 
In both the panels, the scale bar corresponding to 5 pc is shown in the bottom left corner.}
\label{fig8}
\end{figure*}

Finally, to obtain the temperature and column density maps, a modified blackbody was fitted to the observed fluxes on a pixel-by-pixel basis 
\citep[see equations 8 and 9 in][]{mallick15}. 
The fitting was performed using the four data points for each pixel, maintaining the column density ($N(\mathrm H_2)$) and the 
dust temperature (T$_{d}$) as free parameters. 
In the calculations, we adopted a mean molecular weight per hydrogen molecule ($\mu_{H2}$) of 2.8 
\citep{kauffmann08} and an absorption coefficient ($\kappa_\nu$) of 0.1~$(\nu/1000~{\rm GHz})^{\beta}$ cm$^{2}$ g$^{-1}$, 
including a gas-to-dust ratio ($R_t$) of 100, with a dust spectral index ($\beta$) of 2 \citep[see][]{hildebrand83}. 
The temperature and column density maps are shown in Figures~\ref{fig8}a and~\ref{fig8}b, respectively. 

\begin{figure*}
\epsscale{0.53}
\plotone{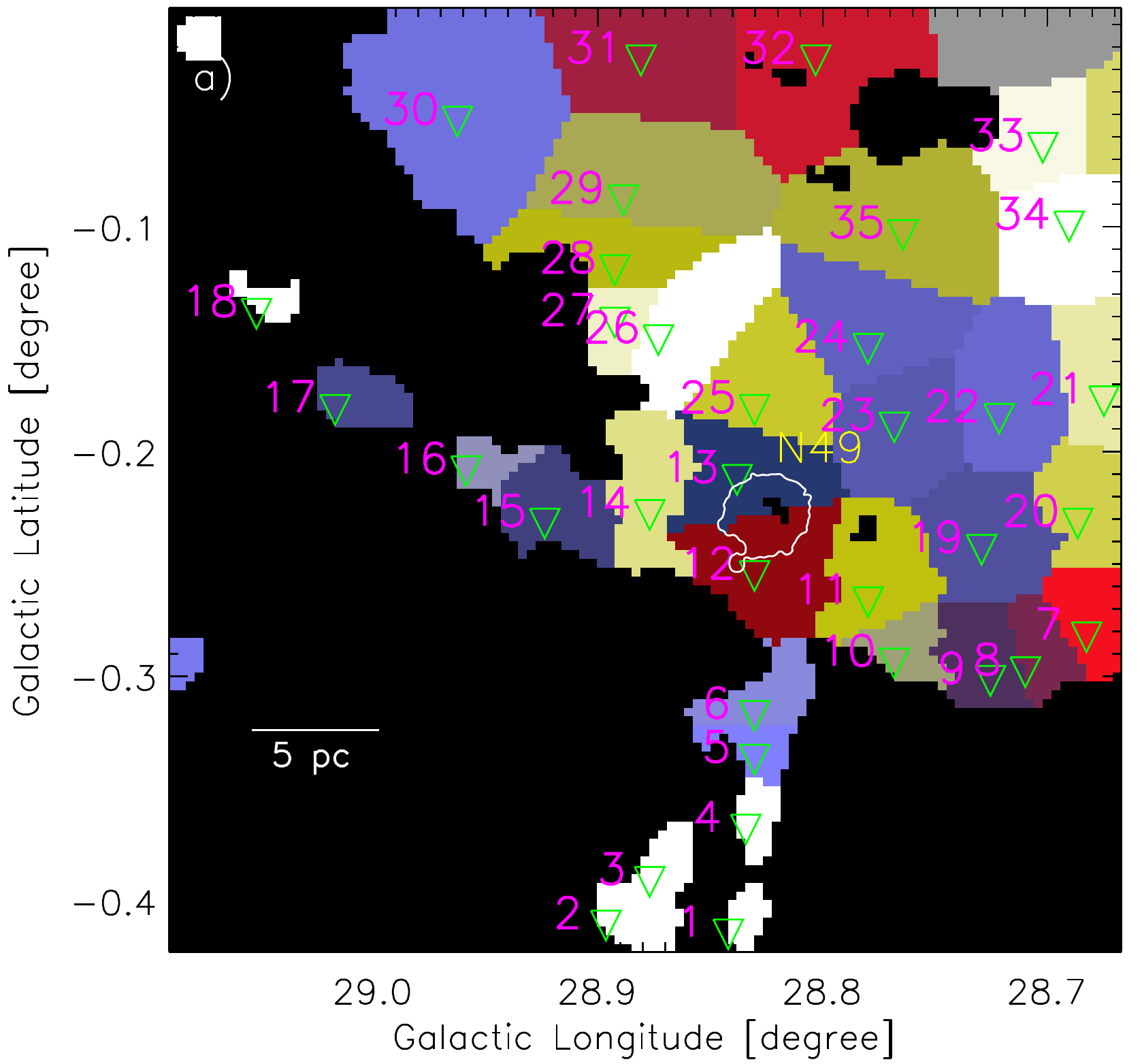}
\epsscale{0.485}
\plotone{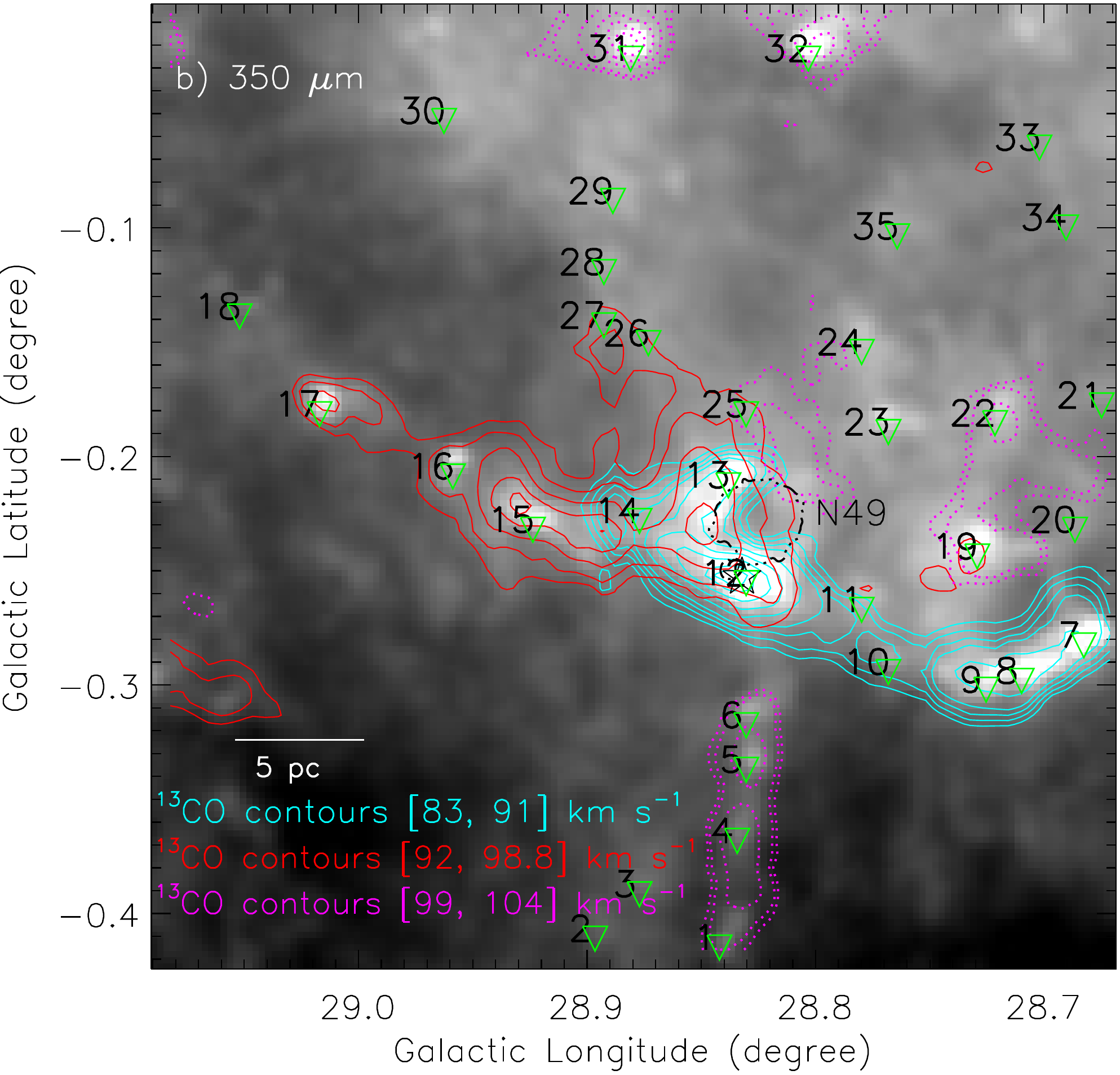}
\epsscale{0.485}
\plotone{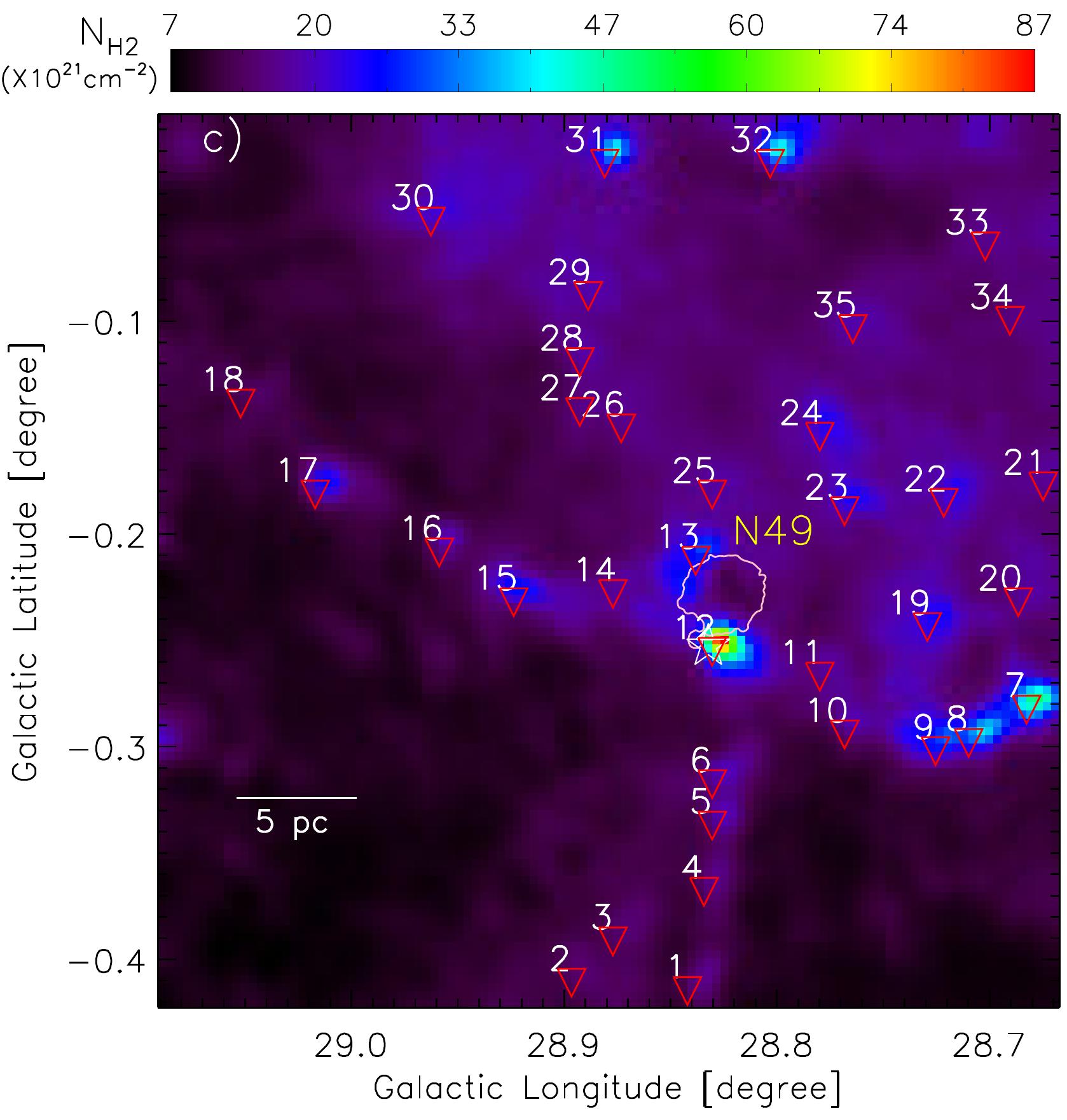}
\caption{\scriptsize The selected {\it Herschel} clumps in our selected target field.
a) The clumps are shown by upside down triangles, and the boundary of each {\it Herschel} clump is also highlighted along with its corresponding clump ID (see Table~\ref{tab1}). A contour at 20 cm (in white) represents the location of the bubble N49, and the contour level is 0.0024 Jy/beam.
b) Overlay of the positions of the clumps (i.e. upside down triangles) along their IDs and the molecular clouds on the {\it Herschel} image at 350 $\mu$m.
The background map and the molecular clouds are similar to those shown in Figure~\ref{fig2xx}. 
c) The positions of the clumps (i.e. upside down triangles) along their IDs are also overlaid on the {\it Herschel} column density map. 
The background map is similar to those shown in Figure~\ref{fig8}b. 
In each panel, the scale bar corresponding to 5 pc is shown in the bottom left corner.}
\label{fig9}
\end{figure*}

The {\it Herschel} temperature map traces the filamentary features in a temperature range of about 16--20~K, 
while the N49 H\,{\sc ii} region is seen with considerably warmer gas (T$_{d}$ $\sim$21-24 K).
The filamentary features and the edges of the bubble N49 are traced in the column density map, where several condensations are observed (see Figure~\ref{fig8}b). 
One can also compute extinction \citep[$A_V=1.07 \times 10^{-21}~N(\mathrm H_2)$;][]{bohlin78} 
using the {\it Herschel} column density map, which can also be used to identify clumps.
In the {\it Herschel} column density map, the ``{\it clumpfind}" \citep{williams94} IDL program helps us to 
find clumps and to estimate their total column densities. 
Thirty five clumps are found in our selected target field, and are highlighted in Figures~\ref{fig9}a,~\ref{fig9}b, and~\ref{fig9}c. Several column density contour levels were used as an input parameter for the ``clumpfind", and the lowest contour level was considered at 3.5$\sigma$. 
Furthermore, the boundary of each clump is also shown in Figure~\ref{fig9}a. 
The knowledge of the total column density of each clump also enables to determine 
the mass of each {\it Herschel} clump using the following equation:
\begin{equation}
M_{clump} = \mu_{H_2} m_H A_{pix} \Sigma N(H_2)
\end{equation}
where $\mu_{H_2}$ is assumed to be 2.8, A$_{pix}$ is the area subtended by one pixel, and 
$\Sigma N(\mathrm H_2)$ is the total column density. 
The mass and the effective radius of each {\it Herschel} clump are listed in Table~\ref{tab1}. 
The clump masses vary between 1076 M$_{\odot}$ and 20970 M$_{\odot}$. 
Three massive clumps (nos. 12, 13, and 14) are also seen in the intersection zone of two molecular clouds (see Figures~\ref{fig9}a and~\ref{fig9}b). 
Furthermore, the clumps (nos. 7, 8, 9, 10, and 11) These {\it Herschel} clump sizes are larger than the ones used by \citet{deharveng10}. 
are found toward the filamentary feature, ``fl-1",  while 
the filamentary feature, ``fl-2" contains clumps (nos. 15, 16, and 17).
The clumps (nos. 1, 4, 5, and 6) are also identified toward the filamentary feature, ``fl-3".

Previously, using the APEX 870 $\mu$m dust continuum data, \citet{deharveng10} computed
the masses of four clumps varying between 190 and 2300 M$_{\odot}$, which are distributed toward the infrared rim of the bubble.
In this paper, we identify two clumps (nos. 12 and 13; M$_{clump}$ $\sim$8480--11538 M$_{\odot}$) around the bubble in the {\it Herschel} column density map (see Figure~\ref{fig9}c), and the masses of these clumps are much higher than the ones 
reported by \citet{deharveng10} (i.e. M$_{Herschel}$ $>$ M$_{APEX}$). 
More recently, \citet{liu17} studied a star-forming region RCW 79 using the {\it Herschel} data and also compared the masses of clumps derived using the {\it Herschel} data and the APEX 870 $\mu$m continuum map. 
They also found M$_{Herschel}$ $>$ M$_{APEX}$, and suggested that there are mass losses in ground-based observations due to the drawback in the data reduction (see \citet{liu17} for more details). 
\subsection{Young stellar populations}
\label{subsec:phot1}
In this section, we identify embedded YSOs using the {\it Spitzer} photometric data at 3.6--24 $\mu$m. 
A brief description of the selection of YSOs is as follows.\\

\begin{figure*}
\epsscale{0.43}
\plotone{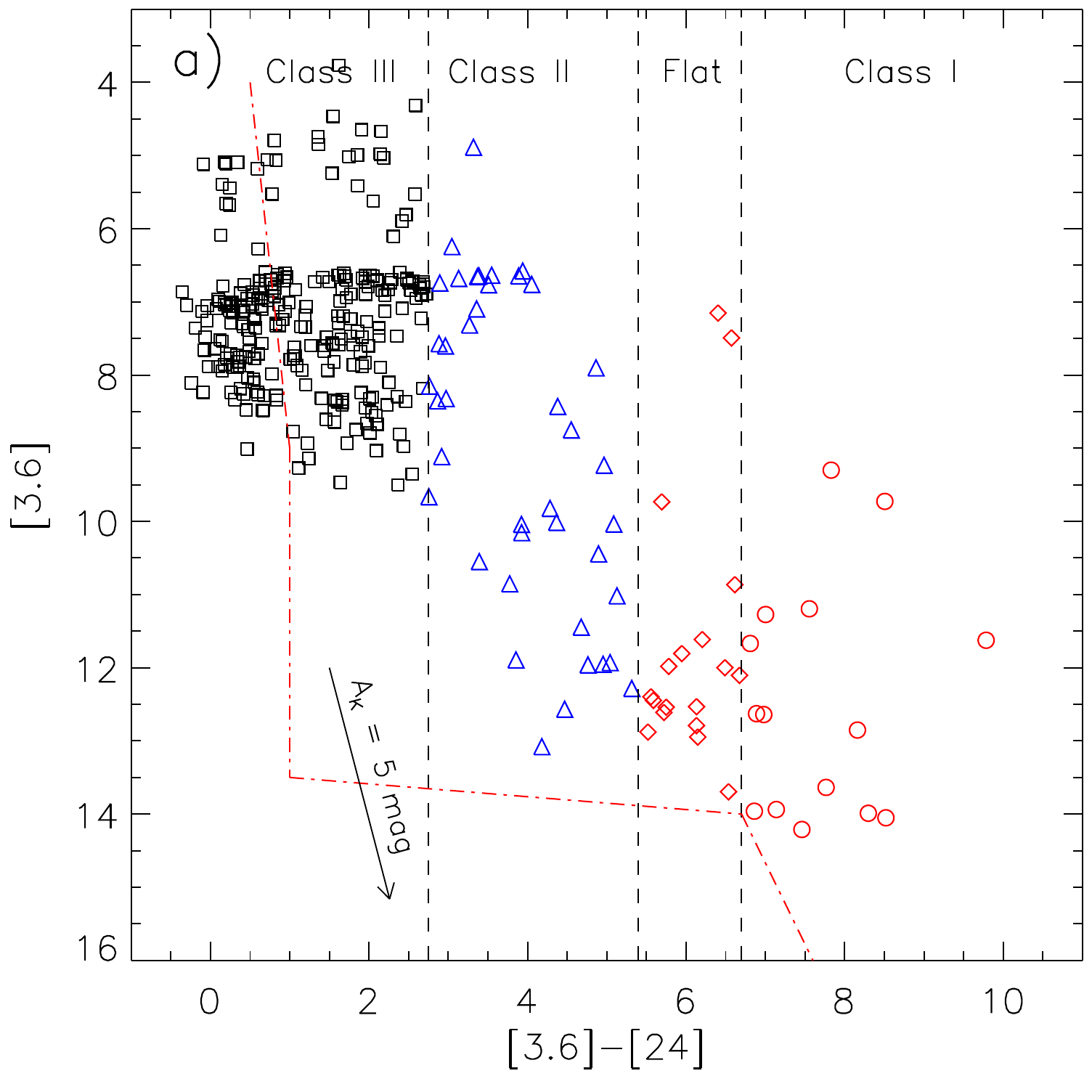}
\epsscale{0.46}
\plotone{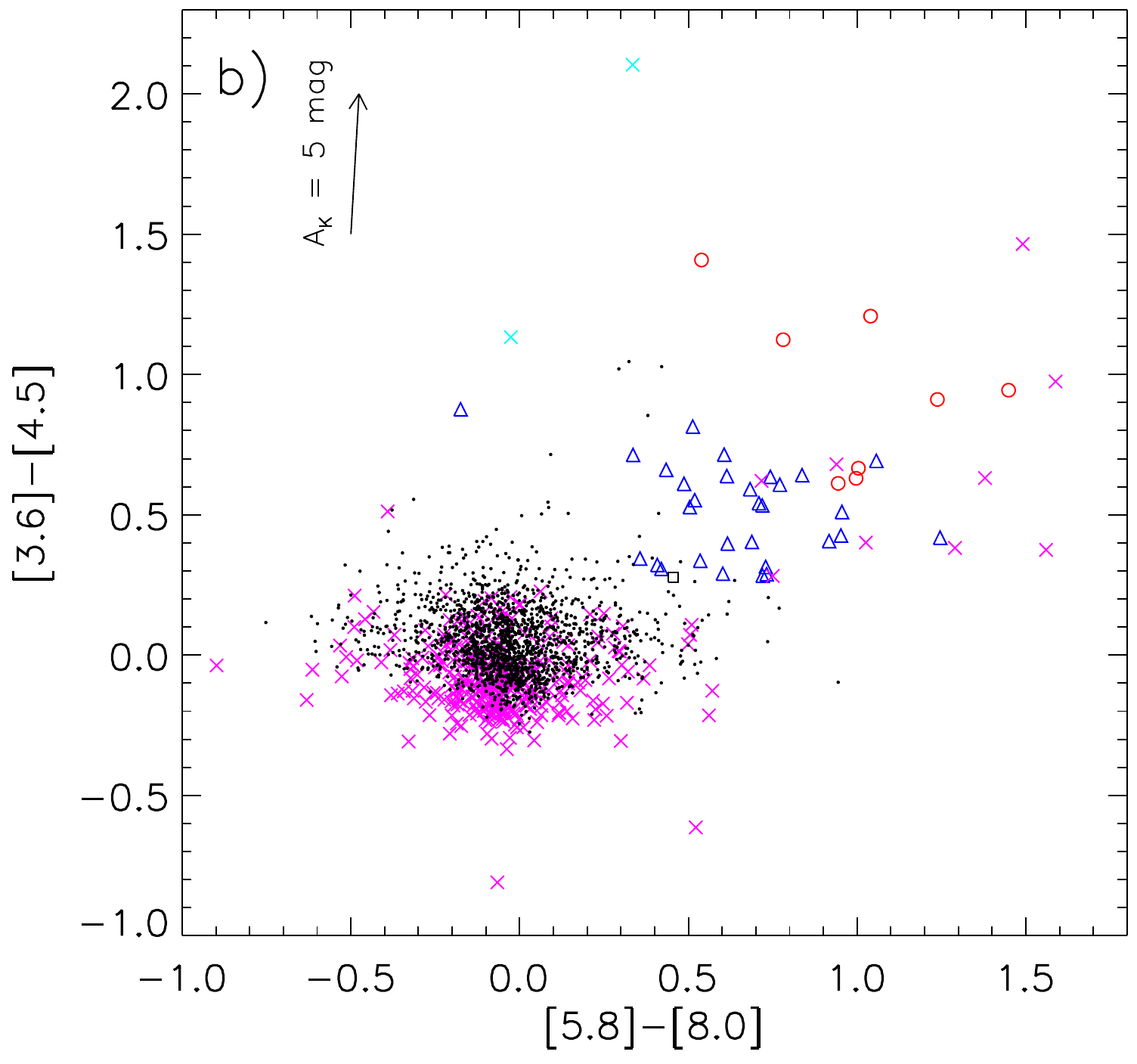}
\epsscale{0.47}
\plotone{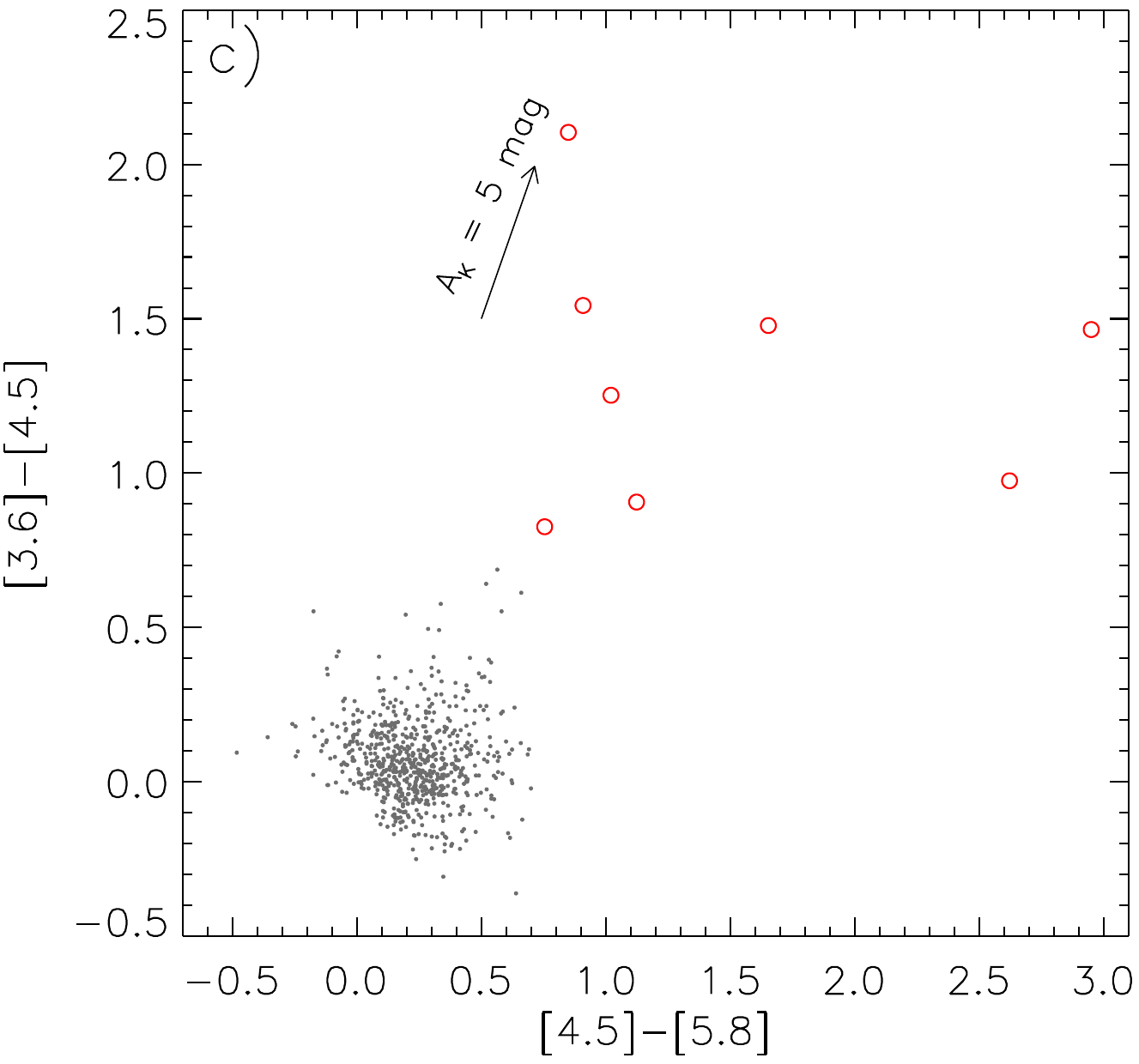}
\caption{\scriptsize Selection of YSOs in our selected field around the MIR bubble N49 (see Figure~\ref{fig1}a). 
a) Color-magnitude plot ([3.6] $-$ [24] vs [3.6]) of sources.
The plot enables to identify YSOs belonging to different evolutionary stages (see dashed lines). 
The boundary of YSOs against contaminated candidates (galaxies and disk-less stars) is shown 
by dotted-dashed lines (in red) (see \citet{rebull11} for more details). 
Flat-spectrum and Class~III sources are represented by ``$\Diamond$'' and ``$\Box$'' symbols, respectively. 
b) Color-color plot ([3.6]$-$[4.5] vs. [5.8]$-$[8.0]) of sources. 
The PAH-emitting galaxies and the PAH-emission-contaminated apertures are marked by ``*'' and ``$\times$'' symbols, respectively (see the text). 
A Class~III source is represented by a black square in the plot. 
c) Color-color plot ([3.6]$-$[4.5] vs. [4.5]$-$[5.8]) of sources. 
In the panels ``a", ``b", and ``c", we show Class~I (circles) and Class~II (open triangles) YSOs.
In the first three panels, an extinction vector is shown and is obtained 
using the average extinction laws from \citet{flaherty07}. 
In the panels ``b" and ``c", the dot symbols show the stars with only photospheric emissions. 
Due to large numbers of stars with photospheric emissions, only some of these stars are randomly marked in 
the panels ``b" and ``c".}
\label{fig10}
\end{figure*}

1. A color-magnitude plot ([3.6]$-$[24]/[3.6]) has been utilized to separate the different stages of YSOs \citep{guieu10,rebull11,dewangan15}. The plot also enables to distinguish the boundary of possible contaminants 
(i.e. galaxies and disk-less stars) against YSOs \citep[see Figure~10 in][]{rebull11}. 
The color-magnitude plot of sources having detections in the 3.6 and 24 $\mu$m bands is shown in Figure~\ref{fig10}a. Adopting the conditions given in \citet{guieu10} and \citet{rebull11}, the boundaries of different stages of YSOs and possible contaminants are highlighted in Figure~\ref{fig10}a.
In Figure~\ref{fig10}a, we have plotted a total of 329 sources in the color-magnitude plot.
We find 74 YSOs (15 Class~I; 18 Flat-spectrum; 41 Class~II) and 255 Class~III sources. 
One can also infer from Figure~\ref{fig10}a that the selected YSOs are free from the contaminants. 
In Figure~\ref{fig10}a, the Class~I, Flat-spectrum, and Class~II YSOs are represented 
by red circles, red diamonds, and blue triangles, respectively.\\  
 
2. Based on the {\it Spitzer} 3.6--8.0 $\mu$m photometric data, \citet{gutermuth09} proposed 
different schemes to identify YSOs and also 
various possible contaminants (e.g. broad-line active galactic nuclei (AGNs), PAH-emitting galaxies, shocked emission 
blobs/knots, and PAH-emission-contaminated apertures). 
One can also classify these selected YSOs into different evolutionary stages based on their 
slopes of the SED ($\alpha_{3.6-8.0}$) estimated from 3.6 to 8.0 $\mu$m 
(i.e., Class~I ($\alpha_{3.6-8.0} > -0.3$), Class~II ($-0.3 > \alpha_{3.6-8.0} > -1.6$), 
and Class~III ($-1.6> \alpha_{3.6-8.0} > -2.56$)) \citep[e.g.,][]{lada06,dewangan11}. 
Following the schemes and conditions 
listed in \citet{gutermuth09} and \citet{lada06}, we have also identified YSOs and various possible contaminants 
in our selected target field. The color-color plot ([3.6]$-$[4.5] vs [5.8]$-$[8.0]) is presented in Figure~\ref{fig10}b. We select 38 YSOs (8 Class~I; 30 Class~II), and 1 Class~III, which are 
plotted in Figure~\ref{fig10}b. In Figure~\ref{fig10}b, Class~I and Class~II YSOs are represented by red circles and blue triangles, respectively.\\  

3. Based on the {\it Spitzer} 3.6, 4.5 and 5.8 $\mu$m photometric data, 
\citet{hartmann05} and \citet{getman07} utilized a color-color plot ([4.5]$-$[5.8] vs [3.6]$-$[4.5]) to 
select embedded YSOs. They proposed color conditions, [4.5]$-$[5.8] $\ge$ 0.7 
and [3.6]$-$[4.5] $\ge$ 0.7, to find protostars. The color-color plot ([4.5]$-$[5.8] vs [3.6]$-$[4.5]) is presented in Figure~\ref{fig10}c. This scheme yields 8 protostars in our selected region.\\

\begin{figure*}
\epsscale{0.58}
\plotone{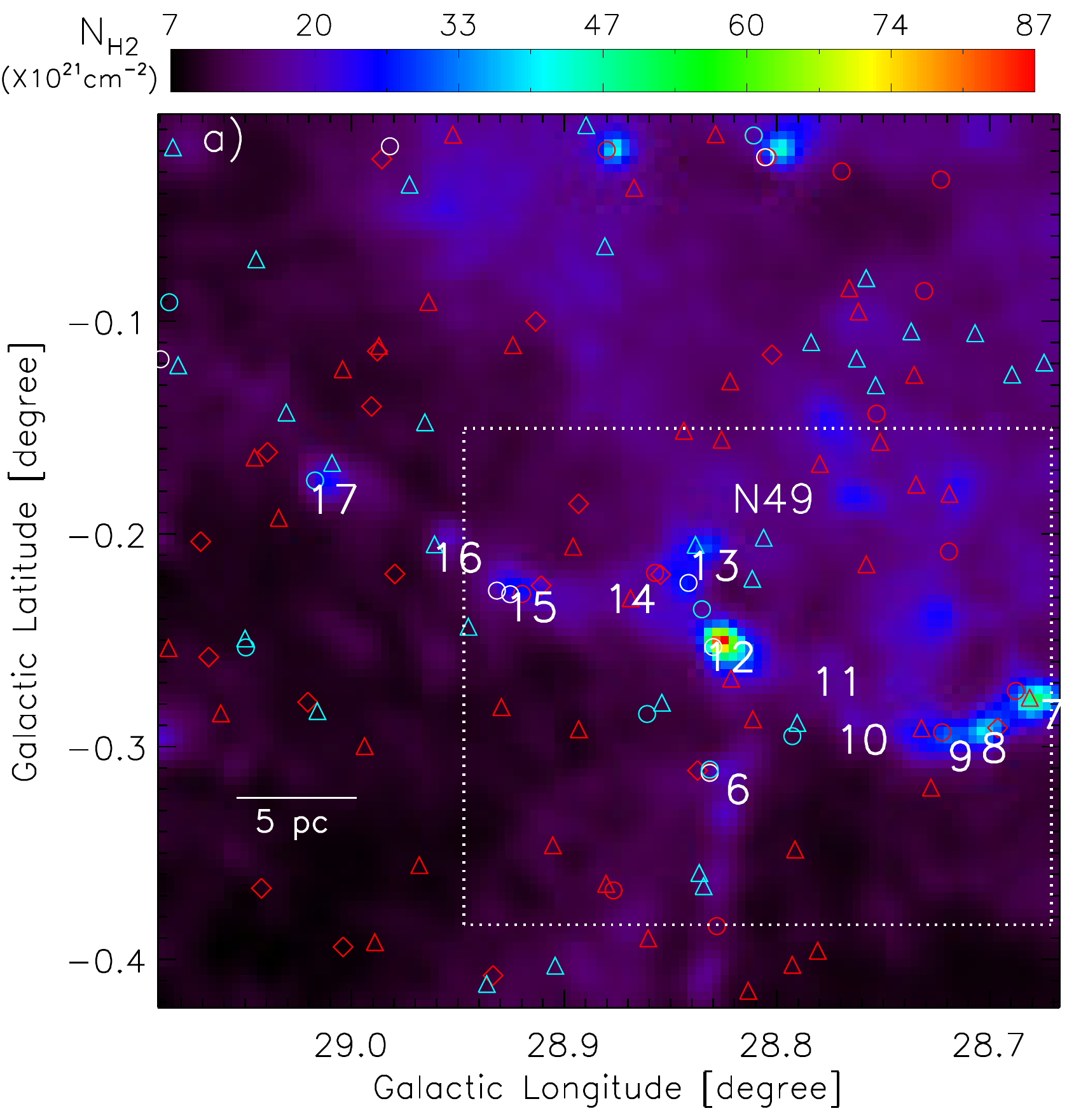}
\epsscale{0.58}
\plotone{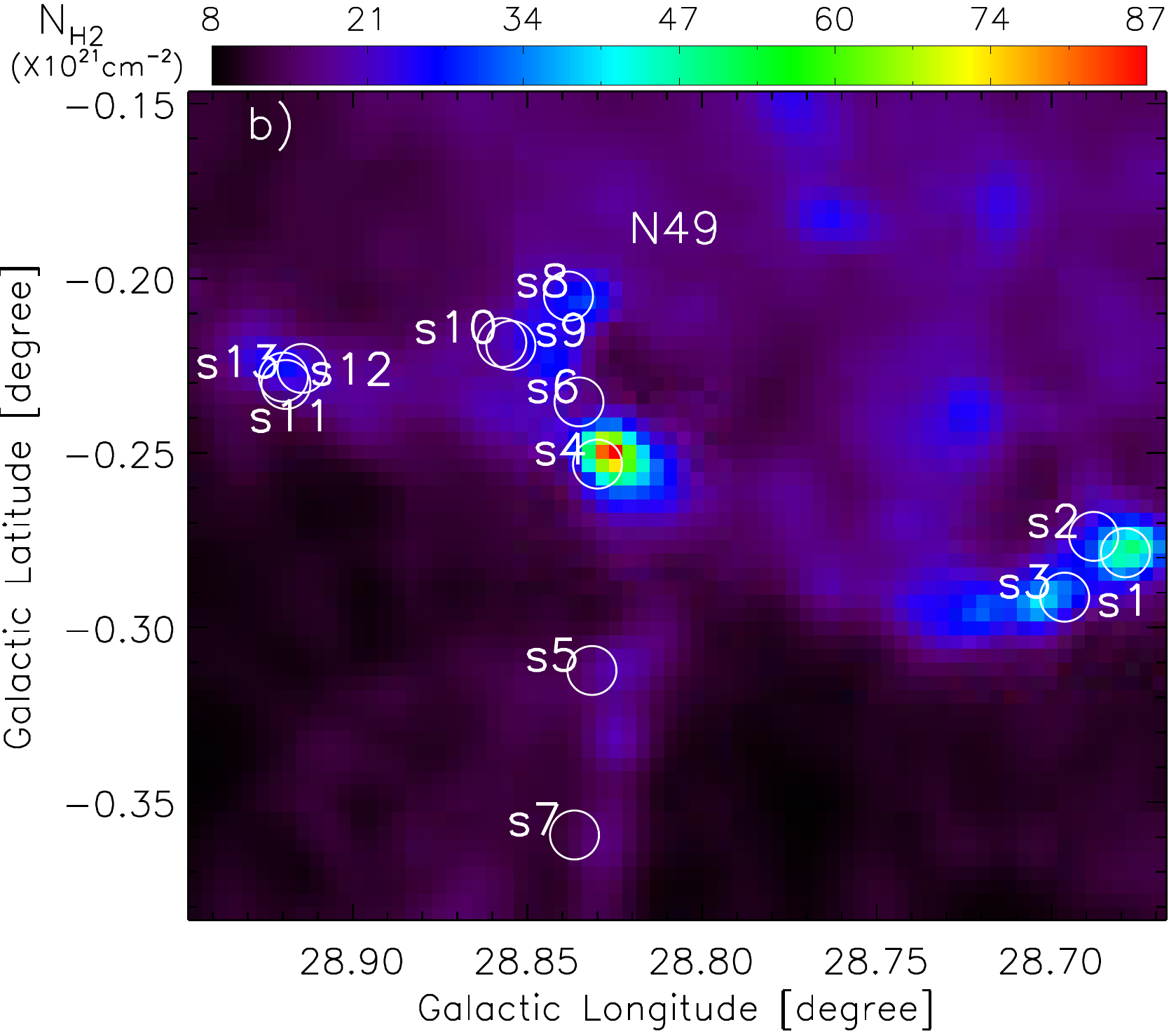}
\caption{\scriptsize a) Spatial distribution of YSOs in our selected field around the MIR bubble N49. 
The YSOs (Class~I (circles), Flat-spectrum (diamond), and Class~II (triangles)) are overlaid on 
the {\it Herschel} column density map.
The background map is similar to the one shown in Figure~\ref{fig8}b. 
The YSOs (in red) are selected using the color-magnitude ([3.6] $-$ [24] vs [3.6]) scheme 
(see Figure~\ref{fig10}a), while the YSOs (in cyan) are identified using the 
color-color ([3.6]$-$[4.5] vs. [5.8]$-$[8.0]) scheme (see Figure~\ref{fig10}b). 
The Class~I YSOs selected via the color-color ([3.6]$-$[4.5] vs. [4.5]$-$[5.8]) 
scheme are highlighted by white circles (see Figure~\ref{fig10}c). 
The {\it Herschel} clump IDs 6--17 are also marked in the figure (also see Figure~\ref{fig9}b). 
A broken box (in white) encompasses the area shown in Figure~\ref{fig11}b. 
b) The positions of some selected YSOs are marked along their IDs on the 
{\it Herschel} column density map (see white circles and also Table~\ref{tab2}). 
These YSOs are found toward the edges of the bubble and the filamentary features. 
The physical properties of these YSOs are listed in Table~\ref{tab2}.}
\label{fig11}
\end{figure*}

Taken together, we have obtained a total of 120 YSOs in our selected target field. 
To examine the spatial distribution of these selected YSOs, in Figure~\ref{fig11}a, 
these YSOs are shown on the {\it Herschel} column density map. 
We find noticeable YSOs toward the filamentary features and the edges of the bubble N49.
It shows signs of an ongoing star formation in the clumps linked with the filamentary 
features (see clump nos. 6--17 in Figure~\ref{fig11}a). 
Previously, using the {\it Spitzer} photometric data, \citet{dirienzo12} carried out the 
SED fitting of sources in the bubble N49 site to identify YSOs. 
The previous results concerning the selection of YSOs are in a good agreement with our presented results. 
Hence, we have taken the physical parameters (e.g. stellar mass (M$_{*}$) and stellar luminosity (L$_{*}$)) of some selected YSOs from \citet{dirienzo12}, 
which are distributed toward the filamentary features (fl-1, fl-2, and fl-3) and the edges of the bubble N49 
(see Figure~\ref{fig11}b and also Table~\ref{tab2}). One can find more details about the SED fitting procedures of YSOs in \citet{dirienzo12}. In Table~\ref{tab2}, we have listed the physical parameters of the selected thirteen YSOs, and their positions are marked in Figure~\ref{fig11}b.
In the direction of filamentary feature fl-1, at least three YSOs (i.e. s1, s2, and s3; M$_{*}$ $\sim$3.5--5.0 M$_{\odot}$) appear to be found toward the {\it Herschel} clumps (nos. 7, 8, and 9; M$_{clump}$ $\sim$3760--5570 M$_{\odot}$). At least three YSOs (i.e. s11, s12, and s13; M$_{*}$ $\sim$0.1--1.6 M$_{\odot}$) are embedded within the {\it Herschel} clumps (nos. 14 and 15; M$_{clump}$ $\sim$5480--5800 M$_{\odot}$) in the direction of filamentary feature fl-2, 
while the filamentary feature fl-3 harboring the {\it Herschel} clumps (nos. 4 and 6; M$_{clump}$ $\sim$1400--3160 M$_{\odot}$) contains at least two YSOs (i.e. s5 and s7; M$_{*}$ $\sim$2.4--4.0 M$_{\odot}$). 
Furthermore, at least five YSOs (i.e. s4, s6, s8, s9, and s10; M$_{*}$ $\sim$1.6--6.2 M$_{\odot}$) 
are found toward the edges of the bubble N49, 
where two {\it Herschel} clumps (nos. 12 and 13; M$_{clump}$ $\sim$8480--11540 M$_{\odot}$) are traced. 
Furthermore, the {\it Herschel} clump 12 also contains a MYSO candidate (i.e. an IRc of the 6.7 GHz MME) at its early formation
stage prior to the UCH\,{\sc ii} phase (see Section~\ref{subsec:u1}). 

Together, low- and intermediate-mass stars are seen toward the filamentary features, 
and various early evolutionary stages of massive star formation (O-type star, UCH\,{\sc ii} 
region, and an IRc of the 6.7 GHz MME without any ionized emission) are also investigated in the bubble site.
\section{Discussion}
\label{sec:disc}
Previously, the bubble N49 has been extensively studied to assess the star formation process 
triggered by the expansion of an H\,{\sc ii} region \citep[see][and references therein]{dirienzo12}. 
However, the present work provides new insights into the physical processes in the MIR bubble N49 site. 
A careful analysis of the large-scale environment of the bubble N49 has been performed in the present work. 
At least three different filamentary features (or filamentary molecular clouds) are identified in our selected target 
field, and the bubble N49 is seen at the interface of the two filamentary molecular clouds (see Section~\ref{subsec:u1} and also Figure~\ref{fig2xx}). 
Several numerical simulations of the CCC process have been carried out and are available in the 
literature \citep[e.g.][]{habe92,anathpindika10,inoue13,takahira14,haworth15a,haworth15b,torii17,bisbas17}. 
More details of these simulations can be found in \citet{torii17} and \citet{dewangan17b}. 
Using the magnetohydrodynamical (MHD) numerical simulations, \citet{inoue13} proposed that 
the colliding molecular gas has ability to form dense and massive cloud cores, precursors of massive stars, in the shock-compressed interface, illustrating a theoretical framework for triggered O-star formation. 
The observational characteristic features of the CCC are reported in the literature, 
which are the complementary distribution of the two colliding clouds, the bridge feature at the intermediate velocity range, and its flattened CO spectrum (e.g. \citet{torii17} and references therein). 
The detection of a broad bridge feature in the velocity space represents an evidence of a compressed layer of gas due to the collision 
between the clouds seen along the line of sight \citep[e.g.,][]{haworth15a,haworth15b,torii17}. 

In our selected target field, a detailed analysis of the molecular line data reveals the bridge feature connecting the two clouds in velocity and the broad CO line wing in the intersection of the two clouds (see Section~\ref{sec:coem}). These evidences confirm that the two clouds are interconnected in space as well as in velocity. The $^{13}$CO profile is obtained in the region ``Reg 1", which is a little away from the H\,{\sc ii} region (see Figure~\ref{fig4}a). 
It is because there are some physical mechanisms (such as radiative/mechanical feedback from massive star) which may destroy the 
observational signatures of the broad bridge feature in the vicinity of the H\,{\sc ii} region(s).

Furthermore, in the velocity channel maps, we also find a possible complementary pair at [87, 88] km s$^{-1}$ and [95, 96] km s$^{-1}$ (see Figure~\ref{fig5}), where the intermediate velocity range between the two clouds was removed.
All these observed signatures are in agreement with the CCC \citep[e.g.,][]{inoue13,takahira14,haworth15a,haworth15b,torii17,dewangan17b}. 
Hence, it appears the onset of the collision of the filamentary molecular clouds in the N49 site. 
The massive clumps, embedded YSOs, an UCH\,{\sc ii} region, and an IRc of the 6.7 GHz MME are also 
observed in the intersection of the two clouds, and are distributed within a scale of $\sim$5 pc. Adopting the velocity separation (i.e. $\sim$7 km s$^{-1}$) of the two clouds, we compute 
a typical collision timescale to be $\sim$0.7 Myr.
A mean age of the Class~I and Class~II YSOs is reported to be $\sim$0.44 Myr and $\sim$1--3 Myr, respectively \citep{evans09}. 
The 6.7 GHz MME also indicates the presence of early phases of massive star formation ($<$ 0.1 Myr). 
These features provide observational evidences to favour the interpretation that the two filamentary 
molecular clouds interacted with each other about 0.7 Myr ago. 
Hence, the birth of massive stars and embedded protostars seen in the interface of the 
clouds appears to be influenced by the CCC process. 
It also implies that one cannot dismiss the possibility of the onset of star 
formation prior to the collision in the bubble site.
\section{Summary and Conclusions}
\label{sec:conc}
In this paper, to study the physical environment and star formation processes, we have carried out 
an observational study of the bubble N49 site using multi-wavelength data.
The major results of the present work are the following:\\
$\bullet$ {\it Herschel} images reveal the bubble N49 and three filamentary features (``fl-1", ``fl-2", and ``fl-3") in our selected target field. In the {\it Herschel} temperature map, the filamentary features are seen in a temperature range of about 16--20~K, 
while the considerably warmer gas (T$_{d}$ $\sim$21-24 K) is found toward the N49 H\,{\sc ii} region.
The filamentary features and the edges of the bubble N49 are traced in the column density map, 
where several condensations are investigated.\\ 
$\bullet$ Using the $^{13}$CO line data, a majority of molecular gas, integrated 
over a velocity range of 83--91 km s$^{-1}$, 
are distributed toward the bubble and the filamentary feature ``fl-1". 
The molecular cloud linked with the filamentary feature ``fl-2" is traced in a velocity 
range of 92--98.8 km s$^{-1}$, while the molecular cloud associated with the 
filamentary feature ``fl-3" is depicted in a velocity range of 99--104 km s$^{-1}$.\\
$\bullet$ The $^{13}$CO line data analysis indicates the presence of two 
velocity cloud components (having velocity peaks at $\sim$88 and $\sim$95 km s$^{-1}$) 
in the direction of the bubble, which are separated by $\sim$7 km s$^{-1}$ in the velocity space 
and are interconnected by a broad bridge feature.\\ 
$\bullet$ In the velocity channel maps of $^{13}$CO, a possible complementary 
molecular pair at [87, 88] km s$^{-1}$ and [95, 96] km s$^{-1}$ is also traced.\\
$\bullet$ The bubble N49 is found in the spatially overlapped zone of two filamentary molecular clouds.\\
$\bullet$ The photometric analysis of point-like sources reveals noticeable YSOs toward 
the filamentary features including the intersection zone of two molecular clouds. 
Different early evolutionary stages of massive star formation (O-type star, UCH\,{\sc ii} 
region, and an IRc of the 6.7 GHz MME without any ionized emission) are also present in the bubble site. \\ 
$\bullet$ A typical collision timescale in the bubble site is computed to be $\sim$0.7 Myr.

Considering the observational outcomes presented in this paper, the bubble N49 is an promising site 
to explore the formation of massive star(s). 
We conclude that in the bubble site, the collision of the filamentary molecular clouds may have 
influenced the formation of massive stars and embedded protostars about 0.7 Myr ago. 
\acknowledgments 
We thank the anonymous reviewer for several useful comments. 
The research work at Physical Research Laboratory is funded by the 
Department of Space, Government of India. 
The Infrared Processing and Analysis Center / California Institute of Technology, 
funded by NASA and NSF), archival data obtained with the {\it Spitzer} 
Space Telescope (operated by the Jet Propulsion Laboratory, California Institute 
of Technology under a contract with NASA). This publication makes use of molecular 
line data from the Boston University-FCRAO Galactic Ring Survey (GRS). 
The GRS is a joint project of Boston University and Five College Radio Astronomy Observatory, 
funded by the National Science Foundation (NSF) under grants AST-9800334, 
AST-0098562, and AST-0100793. The National Radio Astronomy Observatory is a 
facility of the National Science Foundation operated under cooperative 
agreement by Associated Universities, Inc.
IZ is supported by the Russian Foundation for Basic Research (RFBR). 
\begin{deluxetable}{ccccccc}
\tablewidth{0pt} 
\tablecaption{Physical parameters of the {\it Herschel} clumps (see Figures~\ref{fig9}a and~\ref{fig9}b). 
Column~1 gives the IDs assigned to the clump. Table also lists 
Galactic coordinates (l, b), deconvolved effective radius (R$_{clump}$), and clump mass (M$_{clump}$). 
The clumps (nos. 12, 13, and 14) are highlighted by daggers, and are found in the intersection zone of two filamentary molecular clouds (see Figure~\ref{fig9}b). \label{tab1}} 
\tablehead{ \colhead{ID} & \colhead{{\it l}} & \colhead{{\it b}} & \colhead{R$_{clump}$}& \colhead{M$_{clump}$}\\
\colhead{} &  \colhead{[degree]} & \colhead{[degree]} & \colhead{(pc)} &\colhead{($M_\odot$)}}
\startdata 

          1  &   28.842   &  -0.415	&   1.0  &     1076  \\
          2  &   28.897   &  -0.411	&   1.4  &     1821  \\
          3  &   28.877   &  -0.392	&   1.4  &     1954  \\
          4  &   28.834   &  -0.368	&   1.2  &     1397  \\
          5  &   28.830   &  -0.337	&   1.3  &     1836  \\
          6  &   28.830   &  -0.318	&   1.7  &     3161  \\
          7  &   28.683   &  -0.283	&   1.8  &     5317  \\
          8  &   28.710   &  -0.298	&   1.6  &     3761  \\
          9  &   28.725   &  -0.302	&   2.0  &     5569  \\
         10  &   28.768   &  -0.294	&   1.5  &     2638  \\
         11  &   28.780   &  -0.267	&   2.4  &     6177  \\
         12$^{\dagger}$  &   28.830   &  -0.255	&   2.7  &    11538  \\
         13$^{\dagger}$  &   28.838   &  -0.213	&   2.5  &     8480  \\
         14$^{\dagger}$  &   28.877   &  -0.228	&   2.4  &     5800  \\
         15  &   28.924   &  -0.232	&   2.1  &     5483  \\
         16  &   28.959   &  -0.209	&   1.2  &     1599  \\
         17  &   29.017   &  -0.182	&   1.6  &     2876  \\
         18  &   29.052   &  -0.139	&   1.0  &	948  \\
         19  &   28.729   &  -0.244	&   2.7  &     9451  \\
         20  &   28.687   &  -0.232	&   2.1  &     5114  \\
         21  &   28.675   &  -0.178	&   2.1  &     4741  \\
         22  &   28.722   &  -0.185	&   2.8  &     9734  \\
         23  &   28.768   &  -0.189	&   2.7  &     8807  \\
         24  &   28.780   &  -0.154	&   2.9  &    10093  \\
         25  &   28.830   &  -0.182	&   2.6  &     7418  \\
         26  &   28.873   &  -0.150	&   2.7  &     7671  \\
         27  &   28.893   &  -0.143	&   1.5  &     2464  \\
         28  &   28.893   &  -0.119	&   2.3  &     5761  \\
         29  &   28.889   &  -0.088	&   3.4  &    13112  \\
         30  &   28.963   &  -0.053	&   4.2  &    20970  \\
         31  &   28.881   &  -0.026	&   3.2  &    12381  \\
         32  &   28.803   &  -0.026	&   3.3  &    12885  \\
         33  &   28.702   &  -0.065	&   2.2  &     5134  \\
         34  &   28.690   &  -0.100	&   2.5  &     6592  \\
         35  &   28.764   &  -0.104	&   3.5  &    12804  \\
 \enddata  
\end{deluxetable}

\begin{deluxetable}{ccccccc}
\tablewidth{0pt} 
\tablecaption{Physical parameters derived from SED fitting of some selected YSOs (see Figure~\ref{fig11}b), 
which are taken from \citet{dirienzo12} (see Table~4 in their paper). Column~1 gives the IDs assigned to the YSO. 
Table also lists 
YSO designation, $\chi^{2} _{best}$ per data points, stellar mass (M$_{*}$), and stellar luminosity (L$_{*}$) 
(see \citet{dirienzo12} for more details). 
The YSOs distributed toward the edges of the bubble, filaments fl-1, fl-2, and fl-3 
are highlighted with labels ``bub", fl1, fl2, and fl3, respectively (see Figure~\ref{fig11}b). \label{tab2}} 
\tablehead{ \colhead{ID} & \colhead{YSO} & \colhead{($\chi^{2} _{best}$/n$_{data})$} & \colhead{M$_{*}$}& \colhead{L$_{*}$}\\
\colhead{} &  \colhead{(designation ({\it l}{\it b}))} & \colhead{} & \colhead{($M_\odot$)} &\colhead{($L_\odot$)}}
\startdata 

          s1$^{fl1}$  &   028.6788-00.2786  &  0.05 &  3.8$\pm$1.1  &	10$^{2.3}$$\pm$10$^{2.6}$   \\
          s2$^{fl1}$  &   028.6879-00.2739  &  0.31 &  3.6$\pm$1.3  &	10$^{2.0}$$\pm$10$^{2.1}$   \\
          s3$^{fl1}$  &   028.6962-00.2913  &  0.47 &  5.0$\pm$1.6  &	10$^{2.9}$$\pm$10$^{3.1}$   \\
          s4$^{bub}$  &   028.8299-00.2532  &  0.39 &  6.2$\pm$2.0  &	10$^{3.1}$$\pm$10$^{3.2}$  \\
          s5$^{fl3}$  &   028.8315-00.3123  &  0.49 &  2.4$\pm$1.5  &	10$^{2.0}$$\pm$10$^{2.4}$  \\
          s6$^{bub}$  &   028.8352-00.2354  &  0.04 &  4.1$\pm$1.4  &	10$^{2.5}$$\pm$10$^{2.8}$  \\
          s7$^{fl3}$  &   028.8365-00.3594  &  0.17 &  4.0$\pm$1.3  &	10$^{2.5}$$\pm$10$^{2.8}$  \\
          s8$^{bub}$  &   028.8382-00.2051  &  0.15 &  3.5$\pm$1.0  &	10$^{2.2}$$\pm$10$^{2.5}$  \\
          s9$^{bub}$  &   028.8547-00.2192  &  1.80 &  2.8$\pm$1.8  &	10$^{2.3}$$\pm$10$^{2.5}$  \\
         s10$^{bub}$  &   028.8573-00.2184  &  0.54 &  1.6$\pm$1.3  &	10$^{1.8}$$\pm$10$^{2.2}$  \\
         s11$^{fl2}$  &   028.9145-00.2258  &  1.60 &  3.1$\pm$0.8  &	10$^{1.8}$$\pm$10$^{2.0}$ \\
         s12$^{fl2}$  &   028.9191-00.2304  &  0.10 &  1.1$\pm$1.0  &	10$^{1.6}$$\pm$10$^{2.0}$  \\
         s13$^{fl2}$  &   028.9198-00.2283  &  1.39 &  5.0$\pm$1.7  &	10$^{2.9}$$\pm$10$^{3.1}$  \\
 \enddata  
\end{deluxetable}

\end{document}